\documentclass[12pt]{article}
\usepackage[paperheight=29.7cm,paperwidth=21cm,textwidth=15cm]{geometry}
\usepackage{blindtext}
\usepackage{setspace}
\usepackage[english]{babel}
\usepackage[utf8]{inputenc}
\usepackage{amsthm}
\usepackage{color}
\usepackage[natbibapa]{apacite}
\usepackage{xurl}
\usepackage[colorlinks,citecolor=blue,urlcolor=blue, linkcolor=blue, bookmarks=false,hypertexnames=true]{hyperref}
\usepackage{url}
\usepackage{float}
\usepackage{graphicx}
\usepackage{doi}
\usepackage{bm}
\usepackage{hyperref}
\usepackage{amsmath}
\usepackage{amssymb}
\usepackage{mathrsfs}
\usepackage[ruled]{algorithm2e}

\bmdefine{\partition}{c}
\bmdefine{\clustcount}{n}

\bmdefine{\sptree}{t}
\newcommand{\sptreeset}{\mathcal{T}}
\newcommand{\compat}{\prec}

\bmdefine{\bbeta}{\beta}
\bmdefine{\bX}{X}
\newcommand{\pproj}{\diamond}
\bmdefine{\cutset}{cutset}
\newcommand{\bx}{\mathbf{x}}
\bmdefine{\btheta}{\theta}
\newcommand{\lobs}{\mathscr{L}^{obs}}
\newcommand{\lpost}{\mathscr{L}^{post}}
\newcommand{\R}{\mathbb{R}}
\newcommand{\N}{\mathbb{N}}

\newtheorem{theorem}{Theorem}
\newtheorem{lemma}{Lemma}
\newtheorem{proposition}{Proposition}

\newcommand{\pkg}[1]{{\fontseries{b}\selectfont #1}}
\title{A tree is all you need}

\author{Etienne Côme$^{*}$\\
        \small $^{*}$Cosys/Grettia, Gustave-Eiffel University, Noisy Champs, France \\ 
        \small \tt{etienne.come@univ-eiffel.fr}
}

\date{} 

\makeatletter
\def\@maketitle{%
  \newpage
  \null
  \vskip 2em%
  \begin{center}%
  \let \footnote \thanks
    {\LARGE \@title \par}%
    \vskip 1em%
    {\large Bayesian contiguity constrained clustering,\par}
    {\large spanning trees and dendrograms\par}%
    \vskip 1.5em%
    {\large
      \lineskip .5em%
      \begin{tabular}[t]{c}%
        \@author
      \end{tabular}\par}%
    \vskip 1em%
    {\large \@date}%
  \end{center}%
  \par
  \vskip 1.5em}
\makeatother

\begin{document}

\maketitle

\begin{abstract} 
\noindent Clustering is a well-known and studied problem, one of its variants, called contiguity-constrained clustering, accepts as a second input a graph used to encode prior information about cluster structure by means of contiguity constraints \textit{i.e.} clusters must form connected subgraphs of this graph. 
This paper discusses the interest of such a setting and proposes a new way to formalise it in a Bayesian setting, using results on spanning trees to compute exactly a posteriori probabilities of candidate partitions. An algorithmic solution is then investigated to find a maximum a posteriori (MAP) partition and extract a \textit{Bayesian dendrogram} from it. The interest of this last tool, which is reminiscent of the classical output of a simple hierarchical clustering algorithm, is analysed. Finally, the proposed approach is demonstrated with real applications. A reference implementation of this work is available in the \pkg{R} package \pkg{gtclust} that accompanies the paper\footnote{available at \href{http://github.com/comeetie/gtclust} {http://github.com/comeetie/gtclust}}.

\end{abstract}


\section{Context, motivations and related works}

\begin{figure}
\begin{center}
\begin{tabular}{cc}
\includegraphics[width=0.48\textwidth]{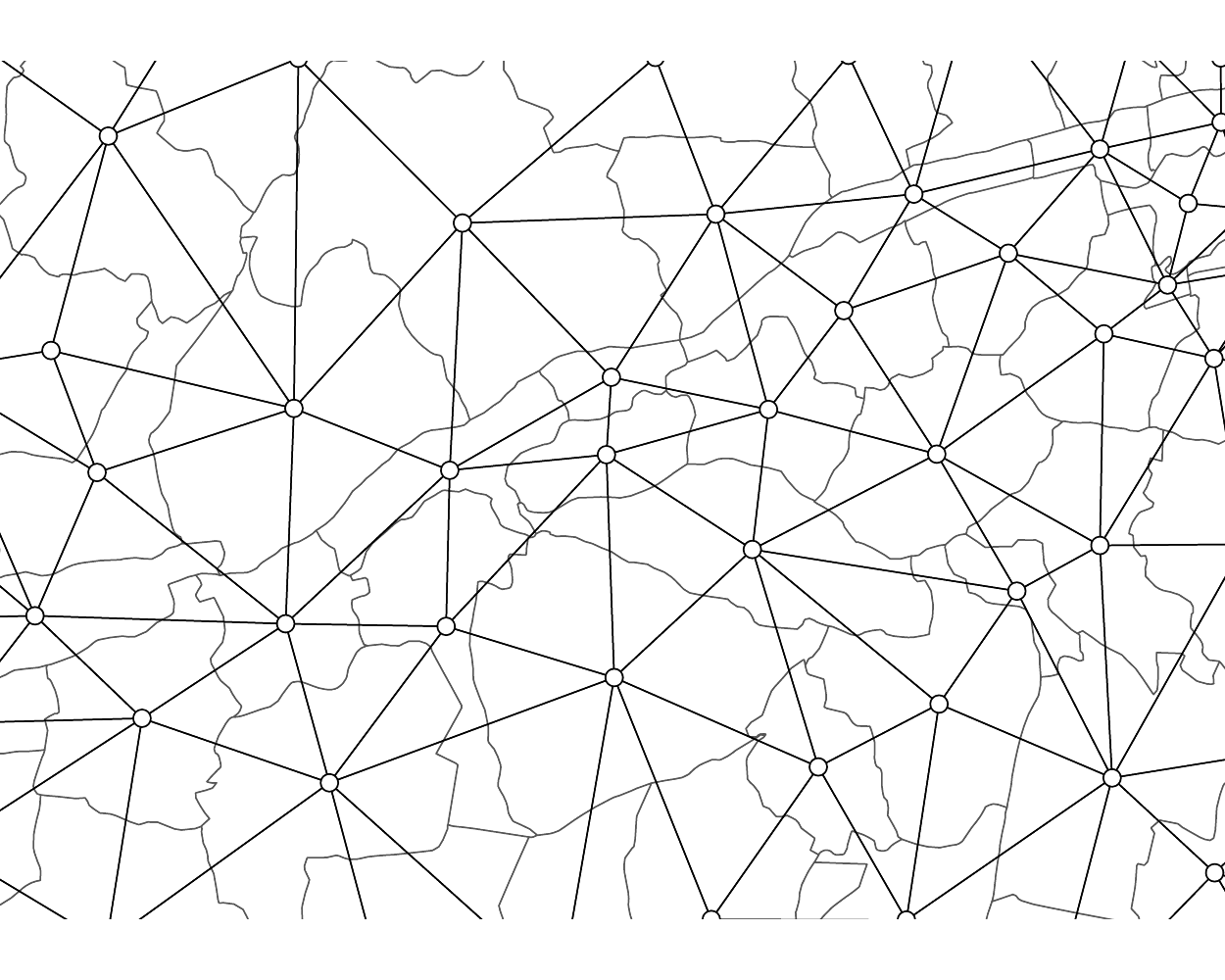}&\includegraphics[width=0.48\textwidth]{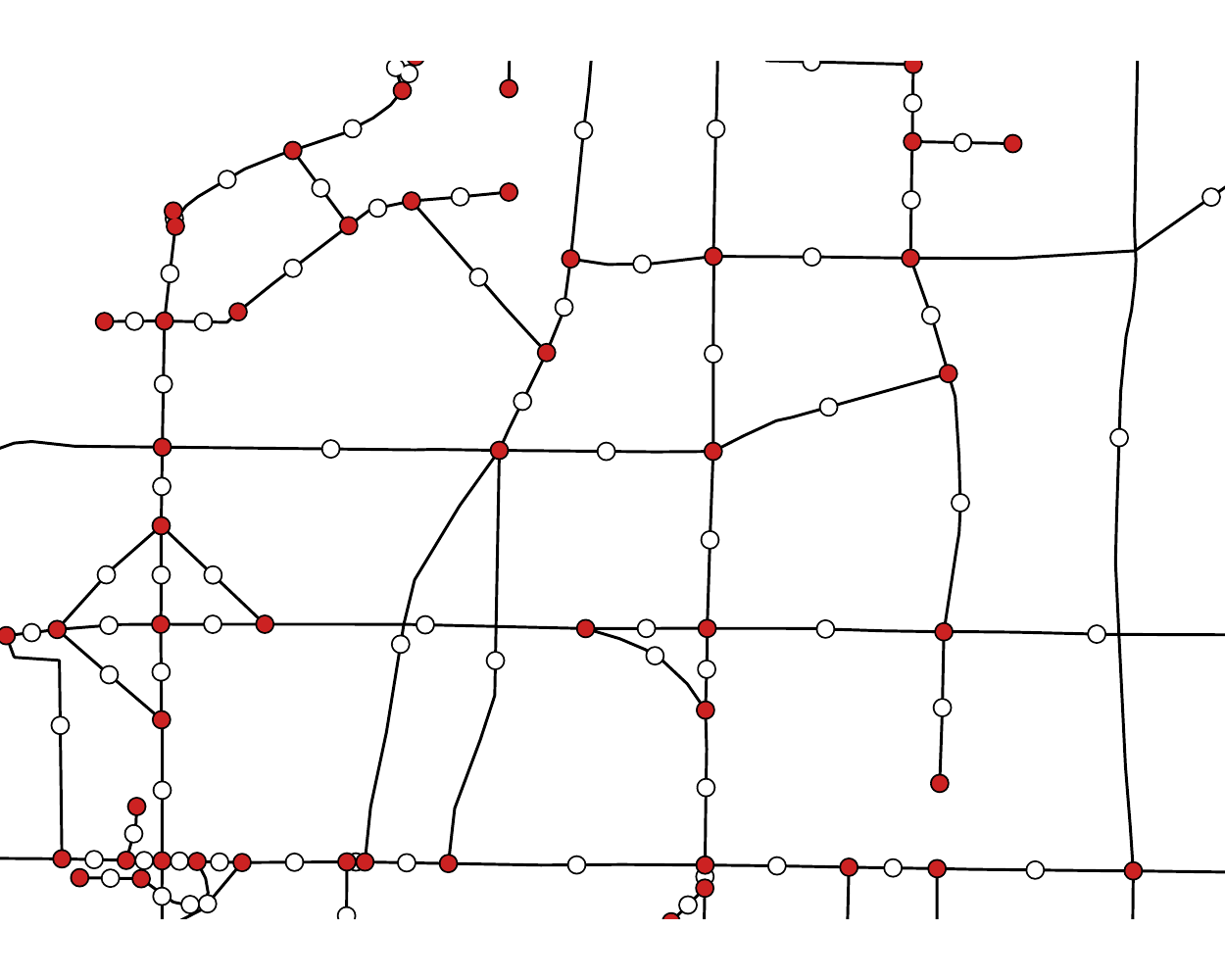}\\
(a)&(b)
\end{tabular}
\caption{\label{fig:nets}\footnotesize{Two input graph examples for contiguity-constrained clustering; (a) neighbourhood graph derived from spatial polygons with shared boundaries (queen adjacency); (b) neighbourhood graph derived from line intersections of a road network: a network between road segments (white dots) is built by connecting any two segments connected to a common road intersection (red dots).}}
\end{center}
\end{figure}

Contiguity-constrained clustering combines two pieces of information: on the one hand, a classical data set $\mathbf{X}$ corresponding to a sample of observations $\{\mathbf{x}_1,\hdots,\mathbf{x}_N\}$ living for example in $\R^d$ or $\N^d$; and on the other hand, an undirected graph $G$ between the same $N$ observations, with edges set $E$. The informal objective being to find a partition $\partition$ of the data points that explains well $\mathbf{X}$ and where each cluster $\partition_k$ forms a \textit{connected sub-graph} of $G$. A subgraph $G[\partition_k]$ is said to be connected if, for any pair of nodes $u$ and $v$ in $\partition_k$, there is at least one \textit{path} connecting them without leaving $\partition_k$. We will note $\partition\compat G$ a partition that fulfils this property. Such a problem allows the use of prior information about the clusters to ensure, for example, that they correspond to spatially coherent regions or time segments. These constraints are quite natural in several applications and greatly reduce the size of the space of possible solutions. Such constraints are therefore interesting from both a computational and an application point of view. Contiguity-constrained clustering has a long history, particularly in geography, where this approach is known as \textit{regionalisation} and has been studied since the seminal papers of \cite{Masser75} and \cite{Openshaw77}. The interest for such a problem in spatial statistics seems natural, since contiguity graphs and distance graphs play an important role in this area \citep{Anselin01}. For example, a contiguity graph can be inferred from spatial polygon data by looking at common boundaries, as shown in Figure~\ref{fig:nets}(a), or from geographic networks (such as road networks) by looking at the intersections between segments, as shown in Figure~\ref{fig:nets}(b). In the former case, contiguity constraints ensure that the clusters found correspond to coherent spatial regions, and in the latter case to connected roads. Since these pioneering contributions, several research papers in the field of quantitative geography have revisited this issue. The Spatial `K`luster Analysis by Tree Edge Removal (SKATER) proposed by \citet{Assuncao2006} is a graph-based method that uses a minimal spanning tree to reduce the search space. The regions are then defined by removing edges from the minimal spanning tree. The removed edges are chosen to minimise a dissimilarity measure. Inspired by SKATER, \citet{Guo2008} proposed REDCAP (Regionalization with Dynamically Constrained Agglomerative clustering and Partitioning), which discusses several variants of agglomerative approaches to solving contiguity constrained clustering problems by varying the affinity metrics and considering full order and partial order constraints.  

Statisticians have also been studying contiguity constrained clustering for some time, and relevant references include the work of \cite{Lebart78}, \cite{Murtagh85}, \cite{Grim87} and \cite{Gordon96}, which study hierarchical agglomerative approaches to solving such a problem. As these references show, the interest in contiguity-constrained clustering and the fact that such constraints can be easily combined with an agglomerative approach has long been known in the statistical community. In addition to applications in geography and statistics, contiguity constraints have also been studied in the context of sequence analysis.   Sequences, of course, can be equipped with simple \textit{line-graph} and therefore benefit from contiguity constraints to solve \textit{segmentation} tasks. This has led to applications of such an approach in genetics to study genome sequences \citep{Ambroise2019} or in time series analysis \citep{Barry93,Schwaller17}. Finally, in the network analysis community, the Louvain \citep{Blondel08} and Leiden algorithms \citep{Traag19}, two of the most well-known graph clustering approaches, leverage, even if not explicitly stated, contiguity constraints to speed up computations and extract \textit{communities}.

In the Bayesian context, a first line of related work deals with the use of Spatial Product Partition Models \citep{Hegarty08,Garrit2016}. PPMs were first introduced by \citet{Hartigan90} and assume that the partition prior $\partition$ is a product of subjective non-negative functions $\kappa(\partition_k)$ called prior cohesions. The cohesion functions measure how likely it is that elements in a given cluster are clustered a priori. This generic setting was then used to define priors that enforce the spatial coherence of the clusters. To do so, \citet{Hegarty08} defines these cohesion functions by counting the number of zones that are neighbours of an element of clusters $\partition_k$ but do not belong to $\partition_k$, while \citet{Garrit2016} considers the locations and spatial distances between the elements of the clusters. Both approaches favour spatially coherent clustering, but do not strictly enforce the contiguity constraints, as we will do in this proposal. This line of research can also be linked to approaches that try to incorporate spatial information into classical clustering approaches, see for example \citep{Chavent2018}, where a Ward-like hierarchical clustering algorithm is extended to minimise, at each stage, a convex combination of a homogeneity criterion computed in feature space and a homogeneity criterion incorporating the spatial dimension of the problem. 

Finally, the approach of \cite{Leonardo2019} is the closest to our proposal since it also relies on spanning trees to define a prior over partitions that strictly enforce contiguity constraints. However, it differs from the present work since it is based on Markov Chain Monte Carlo (MCMC) computations to solve the estimation problem and does not use exactly the same prior/criterion. The interest of our proposal with respect to this approach lies in it's cheaper computational cost and the ease of interpretation of the results (dendrogram and MAP partition). 

As we have just seen, clustering with adjacency constraints is an important problem in unsupervised learning that has been studied for a long time and has important applications in various fields. This paper proposes to revisit this problem and aims at proposing a hierarchical Bayesian approach to solve it. This proposal is based on the definition of a prior on the partition space that enforces the graph-induced contiguity constraints and favours partitions that are easy to disentangle. To this end, the main contributions of the paper are as follows:

\begin{itemize}
\item the definition of a partition prior to ensure the contiguity constraints induced by a graph
\item the derivation of an analytical expression to compute exactly this prior
\item a hierarchical agglomerative algorithm to find an approximate maximum a posteriori partition 
\item a simple approach to building a dendrogram from the computed cluster hierarchy
\end{itemize}

Finally, the open source \pkg{R} package \citep{Rcore} \pkg{gtclust} provides a reference implementation of the algorithm presented in this paper. The implementation is extensible and new models can be integrated. The main computationally intensive methods have been developed in \pkg{Cpp} thanks to the \pkg{Rcpp} package \citep{Eddelbuettel2017}, exploiting the computational efficiency of sparse matrices thanks to the \pkg{Matrix} packages \citep{Eddelbuettel2014,Bates2019} and the \pkg{cholmod} library. The \pkg{c} library \citep{Chen08} for sparse Cholesky matrix factorisation. Finally, the \pkg{sf} package \citep{Pebesma2018} was used to seamlessly integrate the package with the standards used to handle spatial datasets in \pkg{R}.

The remainder of this paper is structured as follows: in Section~\ref{sec:methodology} we present details of our methodology, in Subsection~\ref{subsec:obsmod} we present the mathematical framework we will use and how it can be adapted to different observational models, and in Subsection~\ref{subsec:cc} we discuss the prior we propose and the elements needed to compute it. In Subsection~\ref{subsec:algo} we describe the algorithm we propose to search for an optimal partition, and in Subsection~\ref{subsec:dendo} we show how to build a dendrogram from it. Finally, in Section~\ref{sec:data} we describe several applications of the method to real and simulated data sets. Section~\ref{sec:conclu} concludes the paper. 

\section{Method}
\label{sec:methodology}

Since we will be relying on a non-parametric Bayesian approach, the target we are looking for is an ordered partition of $N$ points with an unknown number of clusters $K$ in $\{1,\hdots,N\}$. More formally, let $\partition \in \mathcal{P}_N$ be an ordered partition of $\{1,\hdots,N\}$, that is:

\begin{align*}
    \partition = \left\{ \partition_1, \ldots, \partition_K \right\} & & \bigcup_k \partition_k = [N] & & \partition_k \bigcap \partition_l = \emptyset & & 
\end{align*}
We assume that all data points belonging to the same element of the partition are $iid$, \textit{i.e.} conditional independence, and we marginalise the cluster parameters so that the probability of the observed data knowing the partition is given by:
\begin{equation}
p(\mathbf{X}\mid \mathbf{c})=\prod_k \int_{\btheta_k} \prod_{i \in c_k}p(\mathbf{x}_i\mid \btheta_k)p(\btheta_k)d\btheta_k,
\end{equation}
where $p(.\mid \btheta_k)$ is a parametric distribution such as a Gaussian that generates the samples coming from group $k$ and $\btheta_k$ its parameter vector. Assuming a clustering objective, we will try to find the MAP partition:
\begin{equation}
\hat{\partition}=\arg\max_{\partition}p(\partition\mid \mathbf{X},G)=\arg\max_{\partition}p(\mathbf{X}\mid \partition)p(\partition\mid G)
\end{equation}
Such a criterion, already successfully used in clustering application \citep{Come21}, corresponds to an exact version of the Integrated Classification Likelihood \citep{Biernacki2000,Biernacki2010} and is thus a penalised criterion. It will therefore allow to automatically tune the number of clusters. To simplify the derivation, we will mainly work with the un-normalised log posterior:
\begin{equation}
\lpost(\partition \mid \mathbf{X},G)=\log\left(p(\mathbf{X}\mid\partition)\right)+\log\left(p(\partition \mid G)\right),
\end{equation}
which has the same maxima. Before discussing the main point of this paper, which concerns the definition of an adequate prior $p(\partition\mid G)$ over the partition to enforce contiguity constraints, we briefly introduce some elements on the computations of the first part of the criterion \textit{i.e.} the probability of the observed data knowing the partition.

\subsection{Exponential family and observation models}
\label{subsec:obsmod}
Computing $\log p(\bX \mid \partition)$ involves computations of quantities $\lobs(\mathbf{X},\partition_k)$:
\begin{equation}
\lobs(\mathbf{X},\partition_k)=\log\left(\int_{\btheta_k} \prod_{i \in c_k}p(\mathbf{x}_i\mid\btheta_k)p(\btheta_k)d\btheta_k\right)
\end{equation}

In this section, we show that such quantities depend on the data $\bX$ only through the sum of sufficient statistics $T(\cdot)$ in exponential family distributions, and that the integral can be computed analytically. To this end, we assume an exponential family observation model for the sample in cluster $k$: 
$$p(\bx\mid\btheta_k)=B(\bx)\exp\left(\Phi(\btheta_k)^\top T(\bx) - A\left( \Phi(\btheta_k)\right) \right),$$
where $\Phi(\btheta)$ is the natural parameter, $A(\cdot)$ is the log partition function and $B(\cdot)$ is a positive normalisation function to ensure that this is a proper density. Then we take the same conjugate prior \citep{diaconis1979conjugate} on $\btheta_k$ for each cluster $k$:
\begin{equation}
    \label{eq:conjugateprior}
    p(\btheta_k \mid \bbeta) = H(\bbeta) \exp \left( \Phi(\btheta_k)^\top \bbeta - A\left(\Phi(\btheta_k)\right) \right),
\end{equation}
where $H$ is another normalisation function. Finally, the integrated observational log-likelihood is obtained via the difference of the normalisation function before and after the update by sufficient statistics: 
\begin{align}
\label{eq:obspost}
    \lobs(\mathbf{X},\partition_k)&=\log\left(\int_{\btheta_k} \prod_{i \in \partition_k} p(\bx_i\mid\btheta_k)p(\btheta_k) d\btheta_k \right),\nonumber\\
    &= \log H \bigg (\bbeta \bigg ) - \log H \bigg (\bbeta + \sum_{i \in \partition_k}T(\bx_i)  \bigg ) + cst.
\end{align}
Thus, any greedy merge heuristics relying on computations of $\lobs(\mathbf{X},\partition_k)$ only has to compute:
\begin{equation}
T_k = \sum_{i \in \partition_k}T(\bx_i),\,\forall k \in \{1,\hdots,K\},
\end{equation}

these will easily fit into a hierarchical approach where there are $N$ evaluations of the $T_i$'s at the beginning and then merging two clusters $g \cup h$ only amounts to summing their respective sufficient statistics $T_{g \cup h} = T_g + T_h$. This property allows the use of a simple \textit{stored-data} approach \cite{Murtagh2012}, where the stored data are the cluster sufficient statistics, to design an agglomerative hierarchical clustering algorithm, and this article will follow this line of work. Note that such an approach could be costly without the contiguity constraints, since all possible merges have to be considered, but that it becomes very competitive in the constrained case, since the constraints remove a lot of candidates, especially when the contiguity graph has a small average degree. Finally, note that a similar approach can be used for swap moves and allows efficient computations for several classical observation models \textit{i.e.} Gaussian, Poisson, Multinomial. We refer the reader to the supplementary material of \cite{Come21} for derivations of specific models. Let us now introduce the main contribution of this work, which concerns the definition of the contiguity-constrained partition prior.

\subsection{Contiguity-constrained partition prior}
\label{subsec:cc}
In the unconstrained case, the classical solutions to define a prior over the space of partitions $p(\partition)$ are mainly based on the Dirichlet process \citep{Rasmussen01} or on uniform priors \citep{Peixoto19}. We will follow a path closer to the latter approach here, and we will start by recognising that a contiguity-constrained prior can be easily defined if the graph $G$ is a tree. In fact, trees have the interesting property of being minimally connected, so removing any edges will create two connected components, and removing $K-1$ edges will create $K$ connected subgraphs. This property shows that there is a one-to-one relationship between combinations of edges and compatible partitions. Thus, if $G$ is a tree, there are $C_{N-1}^{K-1}$ ways of partitioning the vertices into $K$ groups that satisfy the constraints induced by $G$, since removing any $(K-1)$ edges in the tree creates $K$ connected subgraphs and the tree has $N-1$ edges.  Furthermore, there are $K!$ ways of ordering the $K$ clusters, and so we can easily define a uniform distribution for $\partition\mid K,G$ when $G$ is a tree:

$$p(\partition\mid G,K)=\frac{1}{C_{N-1}^{K-1}K!}\mathbf{1}_{\{\partition / \vert \partition \vert = K; \partition \compat G \}}.$$

For the moment, the prior distribution on the number of clusters can also be set to a uniform prior $p(K)=\frac{1}{N}$ to get $p(\partition\mid G)$, but this assumption will be discussed later in Subsection~\ref{subsec:dendo}. The interesting properties of trees that we have just highlighted give us a uniform prior when $G$ is a tree, but do not solve the general case. To do so, following \cite{Leonardo2019}, we will introduce in the setting the set of all \textit{spanning tree} of $G$, denoted by $\sptreeset_G$. A spanning tree $\sptree\in\sptreeset_G$ of a graph $G$ is a connected subgraph with no cycles containing all nodes of $G$. Since a spanning tree is a tree, it inherits the tree properties: any two nodes of $G$ are connected by a unique path in $\sptree$, the number of edges in $\sptree$ is $N-1$, and removing any $(K-1)$ edges divides the vertices of $G$ into $K$ clusters, each cluster being a connected subgraph of $G$. The spanning trees of $G$ can thus be used to define a two-stage prior, as follows:

\begin{enumerate}
\item Sample a spanning tree $\sptree$ of $G$ uniformly: 
$$p(\sptree\mid G) = \frac{1}{| \sptreeset_G|},$$
with $\sptreeset_G$ the set of all spanning tree of $G$.
\item Then remove $K-1$ edges, and draw an ordering for the corresponding clusters: 
$$p(\partition\mid \sptree,K) = \frac{1}{C_{K-1}^{N-1}K!}\mathbf{1}_{\{\partition / \partition\compat\sptree \}}$$
\end{enumerate}

This generative scheme for partitions will produce partitions that are compatible with $G$, since the clusters form connected subgraphs of $\sptree$, and the edges of $\sptree$ are by definition contained in the edges of $G$. To obtain a prior on partition from this process, and since the specific spanning tree used to generate the clustering is not of particular interest, it is natural to marginalise this random variable, which results in the following expression:
\begin{eqnarray}
\label{eq:ccprior}
p(\partition\mid G,K)&=&\sum_{\sptree\in\sptreeset_G}p(\partition\mid \sptree,K)p(\sptree|G)\nonumber\\
&=&\frac{1}{|\sptreeset_G|C_{K-1}^{N-1}K!}\sum_{\sptree\in\sptreeset_G}\mathbf{1}_{\{\partition / \vert \partition \vert = K;  \partition\compat\sptree \}}\nonumber\\
&=&\frac{|\{\sptree / \partition\compat\sptree \}|}{|\sptreeset_G|C_{K-1}^{N-1}K!}  
\end{eqnarray}

At this point, the introduction of spanning trees may seem artificial, but such a prior has interesting properties. This prior is quite informative, since it depends on the constraints defined by $G$, but it goes even further, since it does not correspond to a uniform prior on feasible partitions. In fact, it is known that the probability that an edge $e$ is used in a uniformly sampled spanning tree is a good way of measuring how crucial that edge is for the graph to be connected. This measure, known as \textit{spanning tree centrality}, \citep{Angriman20, Takanori16}, gives us an interesting perspective on the proposed prior. This prior can be seen as a way of extending this measure to partitions of the graph vertices, measuring how easily the different elements of a given partition can be disentangled in the graph. This formalises in an interesting way the kind of prior that practitioners want to express with the contiguity graph. 

Let us now discuss how to analytically compute the quantities involved in this prior. To use such a prior, we need to compute $|\sptreeset_G|$ the number of spanning trees of $G$ and $|\{\sptree / \partition\compat\sptree \}|$ the number of spanning trees of $G$ that are compatible with a given partition $\partition$. These quantities, which at first sight may seem quite difficult to compute, are in fact perfectly amenable to analytical expressions, thanks to the Kirchhoff's theorem; which relates the number of spanning trees of a graph to the spectrum of its Laplacian \cite{Kirchhoff1847,Cayley1889}.

\begin{theorem}[Kirchhoff's theorem]
The number of spanning tree of a graph or multi-graph $G$, without self-loops is given by:
$$|\sptreeset_G|=\frac{1}{N}\lambda_1...\lambda_{N-1},$$
with $\lambda_i$ be the non-zero eigenvalues of the Laplacian matrix $L$ of $G$, given by:
$$L_{ij}= \begin{cases}
        -A_{ij} & \text{if } i\neq j\\
         \sum_{v\in V}A_{iv}& \text{if } i=j
    \end{cases}$$
    where $A$ is the adjacency matrix of $G$, with $A_{ij}$ equal to the multiplicity of edge $(i,j)$ for multi-graphs.
\end{theorem}

Using this theorem, it is therefore possible to compute $\log(|\sptreeset_G|)$. There are several solutions to this, but from a computational point of view it is interesting to note that a consequence of this theorem is that the number of spanning trees is given by any cofactor of $L$, and is thus equal to the determinant of $L$ with one of its rows and columns $(j)$ removed. The Laplacian matrix with row and column $(j)$ removed is denoted by $L_{\text{-}j,\text{-}j}$. To avoid numerical problems, the logarithm of the determinant is used:

\begin{equation}
\log(|\sptreeset_G|) = \log(\det(L_{\text{-}j,\text{-}j}))
\end{equation}

Finally, if the graph is sparse, the computation of the determinant can be solved in reasonable time using a sparse Cholesky decomposition for medium-sized problems. Such a factorisation will give sparse matrices $U$ and $D$ such that $L_{\text{-}j,\text{-}j}=U.D.U^t$ and where $U$ is a lower triangular matrix (with ones on the diagonal) and $D$ is a diagonal matrix and hence $\log(|\sptreeset_G|)=\sum_{i=1}^{N-1}\log(D_{ii})$. This approach solves the problem of computing the number of spanning trees of a given graph, but does not answer the question of the number of spanning trees compatible with a given partition. However, we will see that it can be used as a building block for this. To this end, we can decompose the problem by studying the inter-cluster and intra-cluster connectivity of $G$. The intra-cluster connectivity is easy to analyse by considering the subgraphs given by each element of the partition:

$$G[\partition_k]=\left(\partition_k,\{(u,v) \in E / u \in \partition_k, v\in\partition_k\}\right)$$  

To study the connectivity between clusters, the main objects of interest are the \textit{cutsets} between the different elements of the partition, i.e. the set of edges connecting the different clusters:

$$\cutset(G,\partition_g,\partition_h)=\{(u,v) \in E / u\in \partition_g,v\in\partition_h\},$$

Using these cutsets, we can define an aggregate multi-graph that we will call $G \pproj \partition$, which describes the connection patterns between the clusters:

$$G\pproj \partition = \left(\left\{1,\hdots,|\partition|\right\},\left\{(g,h,|\cutset(G,\partition_g,\partition_h)|),\forall g,h \in \{1,\hdots,|\partition|\}^2\right\}\right),$$
where $(g,h,m)$ represents $m$ edges between $g$ and $h$. This multi-graph thus has as many vertices as there are elements in $\partition$, and the multiplicity of an edge between two vertices $g$ and $h$ is given by the size of the corresponding cutset in $G$. Using this tool, the following proposition can be used to find the number of spanning trees that may have led to a given partition:

\begin{proposition}[Compatible spanning trees]
The number of spanning trees in a graph $G$ compatible with a given partition $\partition$ of its vertices is given by:

$$\log(|\{\sptree / \partition\compat\sptree \}|)=\overbrace{\sum_{k=1}^K\log(|\sptreeset_{G[\partition_k]}|)}^{\text{intra-cluster spanning trees}}+\underbrace{\log(|\sptreeset_{G\pproj\partition}|)}_{\text{inter-clusters spanning trees}}$$
\end{proposition}
\begin{proof}
To form a compatible spanning tree we need to take ($K-1$) edges in the union of the $\cutset(G,\partition_g,\partition_h)$ of each cluster pairs. To get a spanning tree $\sptree$ of $G$ these edges must define a spanning tree of $G\pproj \partition$ , i.e. belongs to $\sptreeset_{G \pproj \partition}$. However, any combination of intra-cluster spanning trees and inter-cluster spanning tree when combined will be compatible with the partition by definition and they will also form a spanning tree of $G$.
\end{proof}

This proposition can then be combined with Kirchhoff's theorem to compute exactly the desired quantity needed to compute the prior defined in Equation~\ref{eq:ccprior}, since the latter formula only involves the number of spanning trees of given graphs, and Kirchhoff's theorem generalises to multi-graphs. So that,

\begin{equation}
\label{eq:lpostK}
\lpost(\partition\mid \bX,G,K)=\sum_{k=1}^K\lobs(\bX,\partition_k)+\log\left(\frac{|\{\sptree / \partition\compat\sptree \}|}{|\sptreeset_G|C_{K-1}^{N-1}K!}\right),  
\end{equation}
can be computed exactly.

\subsection{A greedy agglomerative algorithm}
\label{subsec:algo}
This section deals with the algorithmic details that must be taken into account in order to design an efficient agglomerative algorithm from the previous proposal. In particular, we will see that great care must be taken to avoid unnecessary computations when computing the prior probability of a partition along the merge tree. 

In its simplest form, greedy agglomerative clustering simply reduces to iteratively selecting the best merge action to perform until only one cluster remains. Such a process builds a complete cluster hierarchy, since at each step $t$ of the algorithm with the current partition $\partition^{(t)}$, the number of clusters is reduced by one, and the process starts with each individual data point in its own cluster: 
$$\partition^{(0)}=\left\{\{1\},\{2\},\hdots,\{N\}\right\}.$$
Ideally, we are interested in all partitions $\partition^{(0)},\hdots,\partition^{(N)}$ that maximise $\lpost$, given by the Equation~\ref{eq:lpostK} for $K$ ranging from $N$ to $1$. Such an output can then be used to find the partition $\partition^*$ with maximum a posteriori probability, and to construct a dendrogram as shown in the next section. A greedy agglomerative algorithm with proper merge ranking can be seen as a way to approximate this set of solutions by selecting the best partition $\partition^{(t+1)}$ from the partitions that can be constructed by merging two clusters of $\partition^{(t)}$. Contiguity constraints can, of course, speed up this process by reducing the space of possible merges at each step. More formally, the possible merges compatible with the constraints and a current partition $\partition^{(t)}$ correspond to the support of the multiset of edges of $G\pproj\partition^{(t)}$ that we introduced earlier. At the first iteration, $G\pproj\partition^{(0)}=G$ and therefore the possible merges are defined by the initial graph, so every $(g,h)\in E$ must be inspected and the best one $e^*$ taken. When a merge occurs, the contiguity graph between clusters $G\pproj\partition^{(t)}$ must be updated. This can be done by contracting the edge $e^*=(g,h)$: all edges between $(g,h)$ are removed, as well as their two incident vertices $g$ and $h$, which are merged into a new vertex $u$, where the edges incident to $u$ each correspond to an edge incident to either $g$ or $h$, keeping possible duplicate edges leading to a multi-graph. This ensures that the support of the muti-graph edges always corresponds to all potential merges that fulfil the constraints. We will denote the multi-graph resulting from such an edge contraction by $G/(g,h)$, which allows us to write down the following recurrence relation:

\begin{equation}
G\pproj\partition^{(0)}=G,\,\,
G\pproj\partition^{(t+1)}=G\pproj\partition^{(t)}/e^{(t+1)},
\end{equation}
where $e^{(t+1)}$ is the edge (and thus the merge) selected at step $t+1$ in $G\pproj\partition^{(t)}$. The use of edge contraction thus allows the cluster contiguity graphs $G\pproj\partition$ to be efficiently updated.

To design a greedy agglomerative algorithm for the contiguity constrained problem, the main technical difficulty is to define how to rank the possible merges in an efficient way. Formally, a merge corresponds to modifying the current partition $\partition$, leaving all clusters as they were, except two clusters $g$ and $h$, which are removed and replaced by a single cluster corresponding to their union. We will refer to this merged partition as $\partition^{g\cup h}$:  
$$\partition^{g\cup h}=\left\{\partition_{k, \forall k \neq g,h},\partition_g\cup\partition_h\right\}.$$
To rank the merges compatible with the constraints, we are naturally interested in the effect of this change on the log posterior: 

\begin{equation}
\Delta(g,h)=\lpost(\partition^{g\cup h}\mid \bX,G,K-1)-\lpost(\partition\mid \bX,G,K)
\end{equation}

This quantity can be decomposed into two parts, one coming from the $\lobs$ terms and one from the priors:
\begin{equation}
\label{eq:deltadecomp}
\Delta(g,h)=\Delta{\lobs}(g,h)+\Delta^{prior}(g,h)
\end{equation}

Removing the terms appearing on both the initial partition and the one with the two clusters merged, it is easy to see that:
$$\Delta{\lobs}(g,h)=\lobs(\mathbf{X},\partition_g\cup \partition_h)-\lobs(\mathbf{X},\partition_g)-\lobs(\mathbf{X},\partition_h).$$
This quantity can be easily evaluated using Equation~\ref{eq:obspost} from the cluster sufficient statistics $T_g,T_h$ and $T_{g\cup h}=T_g+T_h$. Furthermore, it depends only on $\partition_g$ and $\partition_h$, and is therefore independent from the other elements of $\partition^{(t)}$. This property is important because we didn't want to update all the possible merge scores $\Delta(g,h)$ at each step, but only the new ones. Following a similar approach for the second term in the right hand side of Equation~\ref{eq:deltadecomp} and simplifying we obtain:

\begin{equation}
\label{eq:deltaprior}
\Delta^{prior}(g,h)=\log\left(\frac{|\sptreeset_{G\pproj\partition/(g,h)}||\sptreeset_{G[\partition_g\cup\partition_h]}|}{|\sptreeset_{G\pproj\partition}||\sptreeset_{G[\partition_h]}||\sptreeset_{G[\partition_g]}|}\right)+\log\left(\frac{K(N-K+2)}{K-1}\right).
\end{equation}

 This second term is more problematic from a computational point of view, because it depends on K and on the whole partition. The terms $|\sptreeset_{G\pproj\partition/(g,h)}|$ and $|\sptreeset_{G\pproj\partition}|$ depend on the whole partition and not only on the clusters $g$ and $h$, and as already explained, this must be avoided. The dependence on $K$ is actually not a problem: at each iteration of the algorithm, the merge to be compared will result in the same number of clusters $K-1$, and therefore the second term on the right and side of Equation~\ref{eq:deltaprior} can be safely ignored to rank the merges. Regarding the dependence on the entire partition, this can also be relaxed by considering a lower bound on $\Delta^{prior}$. In fact, we can use the deletion-contraction recurrence for multigraphs given by Lemma~\ref{lemma:recdelcontr} to get a lower bound on the ratio of the two problematic terms. This bound depends only on the size of the cutset between clusters $g$ and $h$ and is given in Proposition~\ref{prop:deltapriorbound}.   

\begin{proposition}[Lower bound on merge score]\label{prop:deltapriorbound} Given a graph $G$, for all possible partition of its vertices $\partition$, with at least two elements $\partition_g$ and $\partition_h$, the following inequality holds: 
$$\frac{|\sptreeset_{G\pproj\partition/(g,h)}|}{|\sptreeset_{G\pproj\partition}|}\geq \frac{1}{|\cutset(G,\partition_g,\partition_h)|}$$
\end{proposition}
\begin{proof}
This is a direct consequence of applying Lemma~\ref{lemma:recdelcontr} with $G\pproj\partition$ and vertices $g$ and $h$, which gives $|\sptreeset_{G\pproj\partition}|=|\cutset(G,\partition_g,\partition_h)||\sptreeset_{G\pproj\partition/(g,h)}|+|\sptreeset_{G\pproj\partition-(g,h)}|$, and since $|\sptreeset_{G\pproj\partition-(g,h)}|\geq0$ the inequality holds.
\end{proof}
\begin{lemma}[Deletion-contraction recurrence for multi-graphs]
\label{lemma:recdelcontr}
Given a multigraph $G$, without self-loops and more than 3 vertices, for any pair of vertices $g$ and $h$, connected by at least one edge, we have: 
$$|\sptreeset_{G}|=m|\sptreeset_{G/(g,h)}| + |\sptreeset_{G-(g,h)}|,$$
with $m$ the multiplicity of edge $(g,h)$ and $G-(g,h)$ the multigraph with all the $m$ edges between $g$ and $h$ removed. 
\end{lemma}
\begin{proof}
This Lemma is an extension of a classical result on spanning tree deletion-contraction recurrence [ref] to multi-graph. $|\sptreeset_{G}|$ is the sum of the number of spanning trees that use one of the $(g,h)$ edges and the number of spanning trees of $G$ that do not pass through any of the $(g,h)$ edges. The number of spanning trees of $G$ that do not pass through any of the $(g,h)$ edges is given by $|\sptreeset_{G-(g,h)}|$ since every spanning tree of $G-(g,h)$ is a spanning tree of $G$ that do not contains any $(g,h)$ edge and conversely any spanning tree of $G$ that do not contains any $(g,h)$ edge is a spanning tree of $G-(g,h)$. If $\sptree$ is a spanning tree of G containing one of the $(g,h)$ edges, the contraction of $(g,h)$ in both $\sptree$ and $G$ results in a spanning tree $\sptree/(g,h)$ of $G/(g,h)$.
But, if $\sptree^*$ is a spanning tree of $G/(g,h)$, there exists $m$ spanning trees $\sptree$ of $G$, one for each of the $m$ edges between $g$ and $h$, such that $\sptree/(g,h) = \sptree^*$. Thus, the number of spanning trees of $G$ passing trough one the $(g,h)$ edges is $m.|\sptreeset_{G/(g,h)}|$. Hence $|T_{G/(g,h)}|= m.|\sptreeset_{G/(g,h)}|+|\sptreeset_{G-(g,h)}|.$
\end{proof}

Putting the previous elements together, we get the following lower bound for the merge scores, which depends only on the clusters $g$ and $h$:

\begin{equation}
\Delta^{bound}(g,h)=\Delta\lobs(g,h)+\log\left(\frac{|\sptreeset_{G[\partition_g\cup\partition_h]}|}{
|\cutset(G,\partition_g,\partition_h)||\sptreeset_{G[\partition_h]}||\sptreeset_{G[\partition_g]}|}\right)
\end{equation}

This quantity can be computed quite efficiently. During the merging process we keep track of the cluster contiguity graph, updating it with edge contraction, so the size of the cutset between two clusters is readily available. The other terms can also be computed efficiently, noting that we can compute $|\sptreeset_{G[\partition_g\cup\partition_h]}|$ with a \textit{small rank} update of the Cholesky factorisation used to compute $|\sptreeset_{G[\partition_g]}|$ and $|\sptreeset_{G[\partition_g]}|$ since the Laplacian matrix of $G[\partition_g\cup\partition_h]$ can be written as the sum of a block diagonal matrix with the Laplacian matrices of $G[\partition_g]$ and $G[\partition_h]$ on the diagonal plus an update for each edge in the $(g, h)$ cutset:  

\begin{equation}
\label{eq:cholup}
L(G[\partition_g\cup\partition_h])=\left(  \begin{matrix}
    L(G[\partition_g]) & 0 \\
    0 & L(G[\partition_h])  
  \end{matrix}\right)+ 
  \sum_{e\, \in\, \cutset(G,\partition_g,\partition_h)} l_{(e)}l_{(e)}^\top ,
\end{equation}
with $L(G)$ the Laplacian matrix of graph $G$ and $l_{(e)}$, the column vector for edge $e=(u,v)$ with all elements equal to zero, except element $u$ equal to 1 and element $v$ equal to $-1$. This allows the use of efficient numerical routines for updating a sparse Cholesky factorization as the ones provided by the cholmod library \citep{Davis99,Davis01,Davis05}. Finally, once the merging process has stopped, a similar approach can be used, but backwards, to compute the terms $|\sptreeset_{G\pproj\partition^{(K)}}|$ from a previous decomposition of the Laplacian of $\sptreeset_{G\pproj\partition^{(K-1)}}$. This allows us to compute $\lpost(\bX,\partition^{(K)})$ for all the partitions extracted by the hierarchical algorithm, and thus choose the best one. For an overview of the whole process, see Algorithm~\ref{alg:hiera}.

\begin{algorithm*}[ht!]
    \label{alg:hiera}
	\caption{Bayesian Hierarchical Contiguity-constrained clustering}
	\KwIn{$G=(\{1,...,N\},E)$, $\bX$, model $\mathscr{M}$ and prior parameters}
	  Initialize a heap $\mathsf{H}$ with merge costs $\Delta_{g\cup h},\forall (g,h)\in E$, (see Eq~\ref{prop:deltapriorbound})\;
			\While{$K\geq 1$}{
            Pick the best merge $(g,h)^*$ in $\mathsf{H}$ and do it\;
            Store the merge in the merge tree $\mathsf{T}$\;
            Update the number of clusters and the graph by contracting edge $(g,h)^*$\;
            $K = K-1, G = G/(g,h)^*$\;
            Update the exhaustive statistics 
            $T(\partition_g\cup\partition_h)=T(\partition_g)+T(\partition_h)$\;
            Use rank-p Cholesky update to compute \;
            $|\sptreeset_{G[\partition_g\cup\partition_h]}|$ from $|\sptreeset_{G[\partition_h]}|$ and $|\sptreeset_{G[\partition_g]}|$, (see Eq~\ref{eq:cholup})\;
            Compute the new $\Delta$s with the same approach and insert them in $\mathsf{H}$\;    }
			Process the tree backward\;
			\For{$K=2$ to $N$}{
		    Use small rank Cholesky update/downdate to compute\;
            $|\sptreeset_{G\pproj\partition^{(K)}}|$ from $|\sptreeset_{G\pproj\partition^{(K-1)}}|$\;
            Use the result to compute $\lpost(\bX,\partition^{(K)})$, (see Eq~\ref{eq:lpostK})\;
			}
\end{algorithm*}

\subsection{Dendrogram and prior for the number of clusters}
\label{subsec:dendo}
The previous sections have shown how to define a contiguity constrained prior when the number of desired clusters is known, but one of the main interests of the proposed prior is to help choose an appropriate number of clusters. For this purpose, one can use a uniform prior for $K$ over the value $\{1,\hdots,N\}$ and search for a MAP partition for all possible $K$. However, in a clustering application, practitioners may have some prior knowledge about the desired number of clusters and would prefer to gain some insight into the evolution of the model description capabilities with respect to this parameter. A classic tool often used in such a setting is the so-called dendrogram, which represents the merge tree of a hierarchical agglomerative clustering algorithm, with branch heights proportional to the difference in criterion values induced by the corresponding merge. A dendrogram is therefore a binary tree in which each node corresponds to a cluster. The edges connect the two clusters (nodes) merged in a given step of the algorithm. The height of the leaves is generally assumed to be 0. The leaves are ordered by a permutation of the initial clusters that ensures that successive merges are neighbours in the dendrogram. The height of the node corresponding to the cluster created at merge $t$, $h_t$, is often the value of the linkage criterion. 

We propose to revisit this classical tool in the Bayesian context on which this paper is based. To do so, we start by replacing the uniform distribution on the number of clusters $K$ by an informative prior. Considering that the expected value of this random quantity is known, the maximum entropy prior over $\{1,\hdots,N\}$ is given by the truncated geometric distribution:

\begin{equation}
\label{eq:truncgeom}
p(K|\alpha)=\begin{cases}\frac{1}{1-\alpha^N}\alpha^{K-1}(1-\alpha)& \text{if }\alpha\in[0,1[\\
\frac{1}{N}&\text{if }\alpha=1
\end{cases}
\end{equation}

\begin{figure}[ht!]
\begin{center}
\begin{tabular}{cc}
\includegraphics[width=0.45\textwidth]{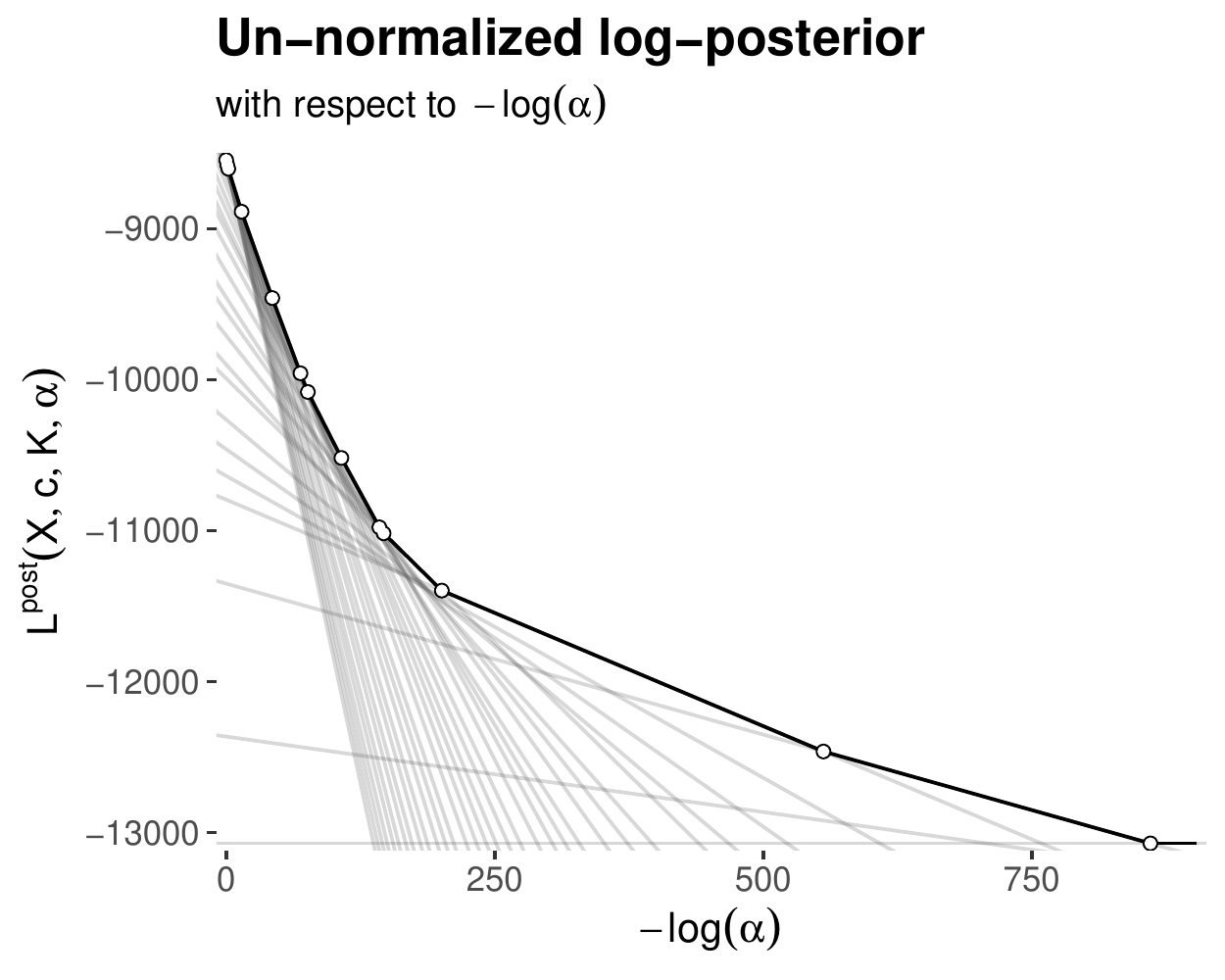}&\includegraphics[width=0.45\textwidth]{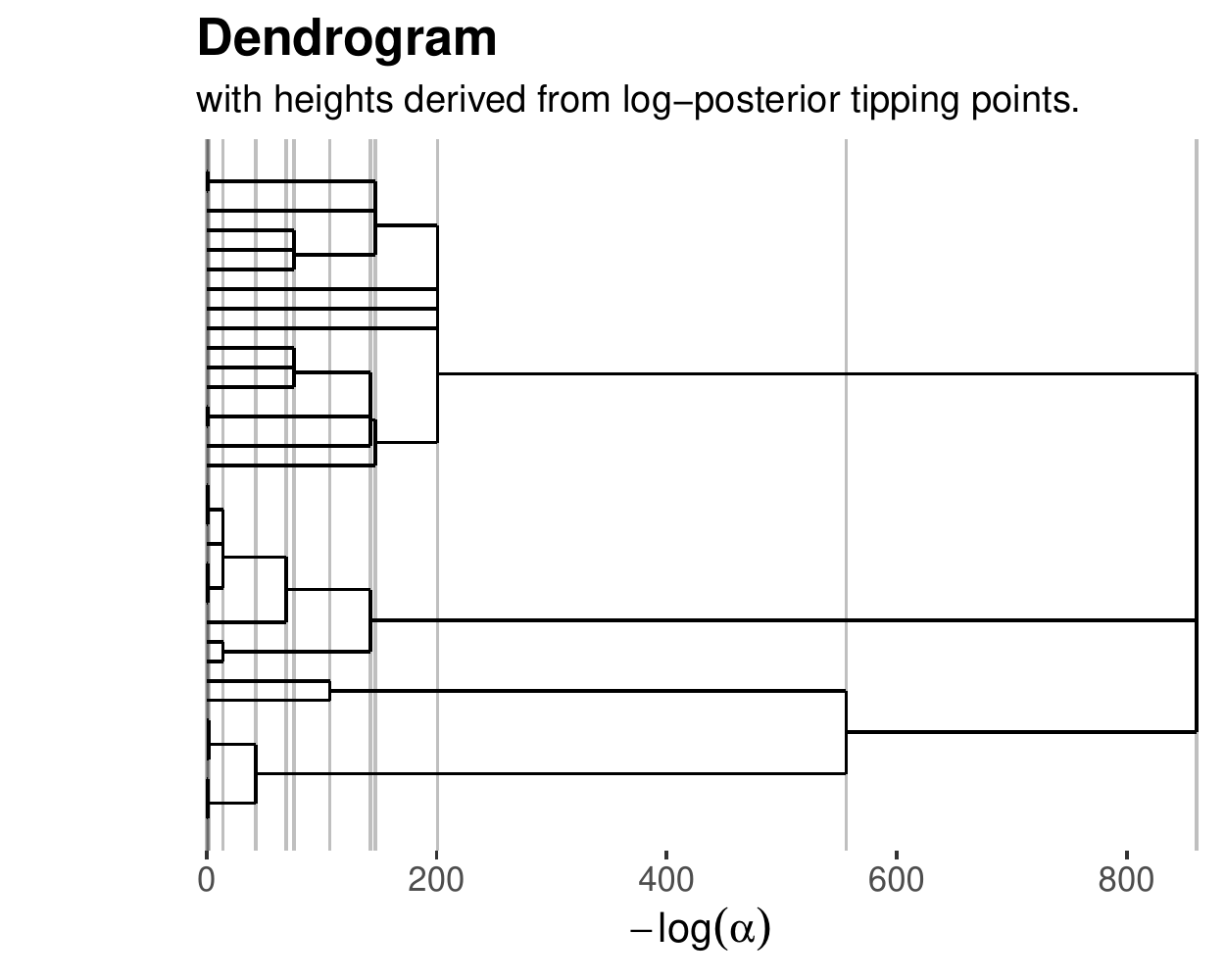}\\
(a)&(b)
\end{tabular}
\caption{\label{fig:dendoex}\footnotesize{Example of posterior front (a) and corresponding dendrogram (b).}}
\end{center}
\end{figure}

Such a prior is therefore a natural way of providing information about the number of clusters. This prior also has the attractive property of including the uniform distribution as a special case. In fact, the prior parameter $\alpha$ allows a smooth interpolation between the extreme prior cases, when $\alpha$ is equal to $1$ a uniform prior is recovered, and when $\alpha$ is equal to $0$ this prior is equivalent to a Dirac distribution at $K=1$. Using a small value of $\alpha$ will therefore lead to solutions with fewer clusters. Thus, $\alpha$ can be seen as a regularisation parameter that gives access to simpler, coarser solutions.  Adapting the approach previously proposed in \citep{Come21} to the non-parametric Bayesian setting of the current proposal, we will show that using this prior together with a greedy agglomerative algorithm that produces a hierarchy of nested partitions allows the exact computation of the sequence of regularisation parameters that unlock the fusions. A difference with the previous proposal is that here the sequence of $\alpha$ values that enable the fusion is computed in an exact manner, without relying on any approximation of the posterior. Going back to the definition of the un-normalised posterior given by Equation~\ref{eq:lpostK}, combining it with the prior defined in Equation~\ref{eq:truncgeom} and making its dependence on the chosen $\alpha$ value explicit, we have:
\begin{equation}
\lpost(\partition|\bX,G,\alpha)=\log\left(p(\bX|\partition)\right)+\log\left(p(\partition|G,K)\right)+\log\left(p(K|\alpha)\right)
\end{equation}
The first two terms of this equation do not depend on $\alpha$ and can be aggregated into $I(\partition,\bX,G)$, then substituting $p(K|\alpha)$ by its value given by Equation~\ref{eq:truncgeom}, we obtain:
\begin{equation}
\label{eq:loglinearpost}
\lpost(\partition|\bX,G,\alpha)=I(\partition,\bX,G)+(K-1).\log(\alpha)+\log\left(\frac{1-\alpha}{1-\alpha^N}\right)
\end{equation}
The last term of Equation~\ref{eq:loglinearpost} does not depend on the size of the solutions (i.e. their number of clusters $K$) and can therefore be ignored when comparing partitions. This shows that it is sufficient to compute the intersection of two lines to get the exact $\alpha$ value where one partition outperforms the other. In fact, we can examine the front of all solutions extracted by a greedy agglomerative algorithm in the $\left(-\log(\alpha),\lpost(\partition|\bX,G,\alpha)\right)$ plane as shown in Figure~\ref{fig:dendoex}~(a) and extract the sequence of regularisation parameters that unlocks the fusions.

These tipping points $-log(\alpha^{(t)})$ can then be used to draw a dendrogram, as shown in Figure~\ref{fig:dendoex}~(a). An important point to note is that some of the partitions extracted by an agglomerative greedy algorithm may not be dominant over the others anywhere in $\alpha\in]0,1]$. This corresponds to situations where combining several merges in one step is better than performing them sequentially. This is quite natural, since $\lpost$ is a penalised criterion, so it does not necessarily increase with model complexity. Since such partitions cannot belong to the approximated Pareto front over $\alpha\in[0,1]$, it is sufficient to remove them and to record only the point corresponding to the real change of the partition on the front; with such an approach, several merges may appear at the same level in the dendrogram, as can be seen by looking carefully at Figure~\ref{fig:dendoex}~(b). This approach offers an interesting solution to the problem encountered with classical dendrograms when working with contiguity constraints. Indeed, when using contiguity constraints, it is not guaranteed that the classical criterion (\textit{i.e.} loss of information) increases \citep{Randriamihamison20}, leading to difficulties (\textit{reversal}) which are avoided here. In conclusion, this approach, although reminiscent of classical dendrograms, has several advantages: it naturally avoids the overplotting problems encountered at the bottom of classical dendrograms, since it focuses only on the interesting part of the merging process; it solves the possible reversal problem. Finally, it provides a natural way to interpret the heights of the merges in the dendrograms as the prior parameter value needed to accept the merge.

\section{Results and discussion}
\label{sec:data}

\subsection{Simulation study}

To study the performance of the algorithm, we first compare its performance on simulated data in a controlled setting with state-of-the-art clustering algorithms with and without adjacency constraints. The data were generated on a regular grid that can be assimilated to an image of 30 by 30 pixels. These images were then divided into 9 square regions (i.e. clusters) of equal size (10 pixels by 10 pixels), each with a different mean, given by:
$$
  \left[ {\begin{array}{ccc}
    1 & 5 & 3 \\
    2 & 7 & 4 \\
    6 & 9 & 8 \\
  \end{array} } \right]
$$

\begin{figure}[ht!]
\begin{center}
\includegraphics[width=0.9\textwidth]{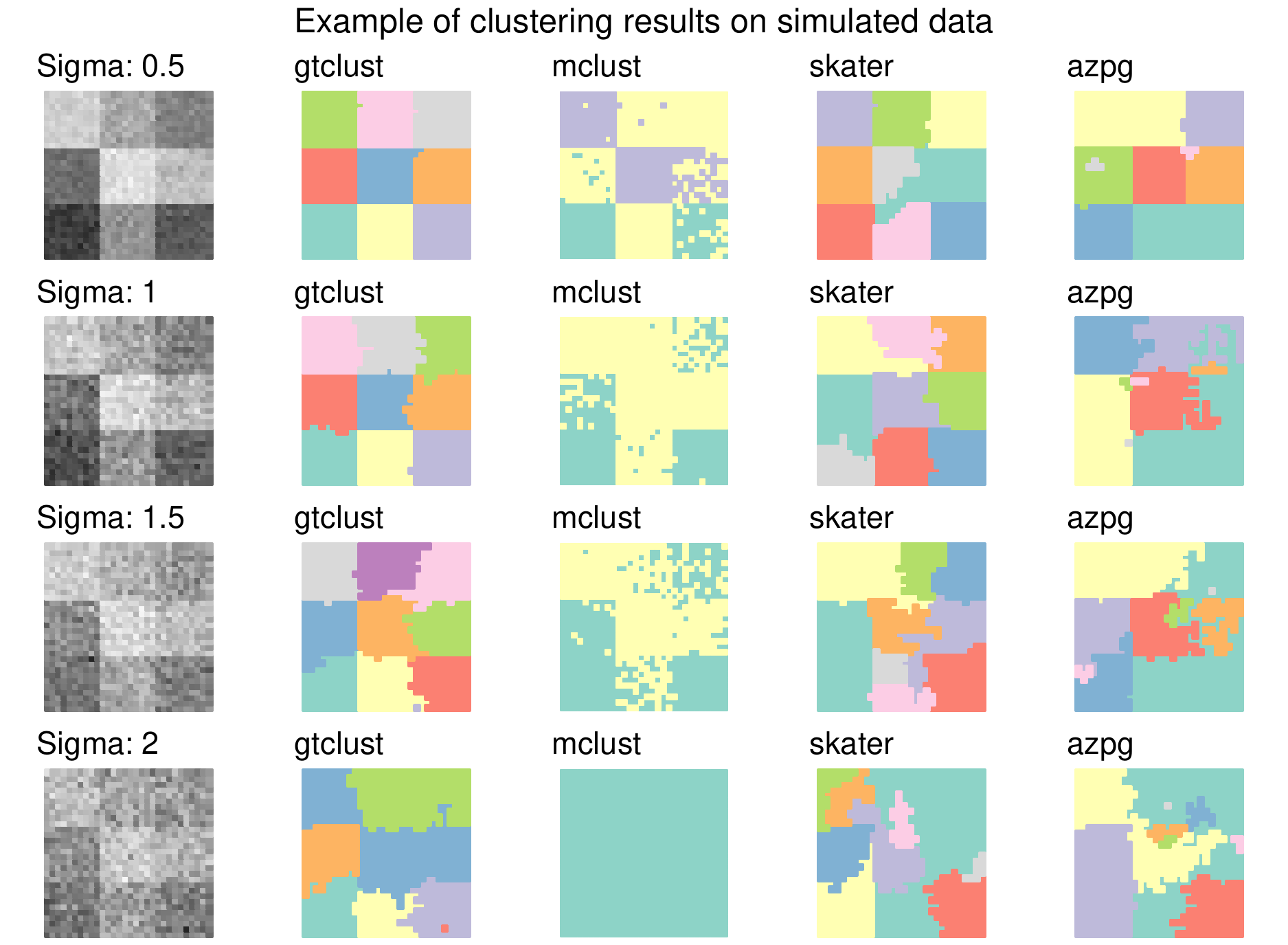}
\caption{\label{fig:exresgrid}\footnotesize{Some visual results of the clustering found on the simulated data, the first column corresponds to the simulated dataset, other columns to the solutions found by \textbf{gtclust}, \textbf{mclust}, \textbf{skater} and \textbf{azpg}. Each row correspond to a different value of $\sigma \in \{0.5,1,1.5,2\}$.}} 
\end{center}
\end{figure}

The data were then simulated with a Gaussian distribution whose mean is determined by the region to which the pixel belongs and whose variance $\sigma$ varied between $\{0. 25,0.5,0.75,1,1.25,1.5,1.75,2,2.25\}$. These simulations allow us to study the performance of the different algorithms in simple situations ($\sigma=0.25,0.5$) as well as in more difficult situations ($\sigma=1.75,2,2.25$). The first column of Figure~\ref{fig:exresgrid} shows a random sample of simulated images for different values of $\sigma$.

We have compared our proposal with different state-of-the-art approaches:

\begin{description}
\item[mclust:] A Gaussian mixture model without contiguity constraints, with automatic model selection and the number of cluster varied between 1 and 20. The implementation of the \pkg{R} package \pkg{mclust}, \citep{Scrucca16} was used to derive those results. 
\item [mclust (9):] The same implementation and algorithm with the number of clusters fixed to the true number of cluster.
\item[skater] The skater algorithm with a number of clusters equal to the true number of cluster. The implementation used is the one available in the \pkg{rgeoda} package. 
\item[redcap:] The redcap algorithm with a number of clusters equal to the true number of cluster. The implementation used is the one available in the \pkg{rgeoda} package. 
\item[azpg:] The azp algorithm with greedy optimization and a number of clusters equal to the true number of cluster. The implementation used is the one available in the \pkg{rgeoda} package. 
\item[azpsa:] The azp algorithm with simulated-annealing optimization and a number of clusters equal to the true number of cluster. The implementation used is the one available in the \pkg{rgeoda} package. 
\end{description}

The solutions obtained with these algorithms were compared with those obtained with our implementation of the algorithm introduced in this paper and available in the \pkg{gtclust} \pkg{R} package with a classical Gaussian model (Gaussian prior for the mean and Gamma prior for the variances, with values $\tau = 0.01$, $\kappa = 1$, $\beta = 0.1s^2$, $\mu = \bar{x}$). We extracted for the analysis two solutions from the hierarchy :
\begin{description}
\item [gtclust:] the MAP partition found by the algorithm.
\item [gtclust (9):] the partition with 9 clutsers found by the algorithm.
\end{description}

\begin{figure}[ht!]
\begin{center}
\includegraphics[width=0.95\textwidth]{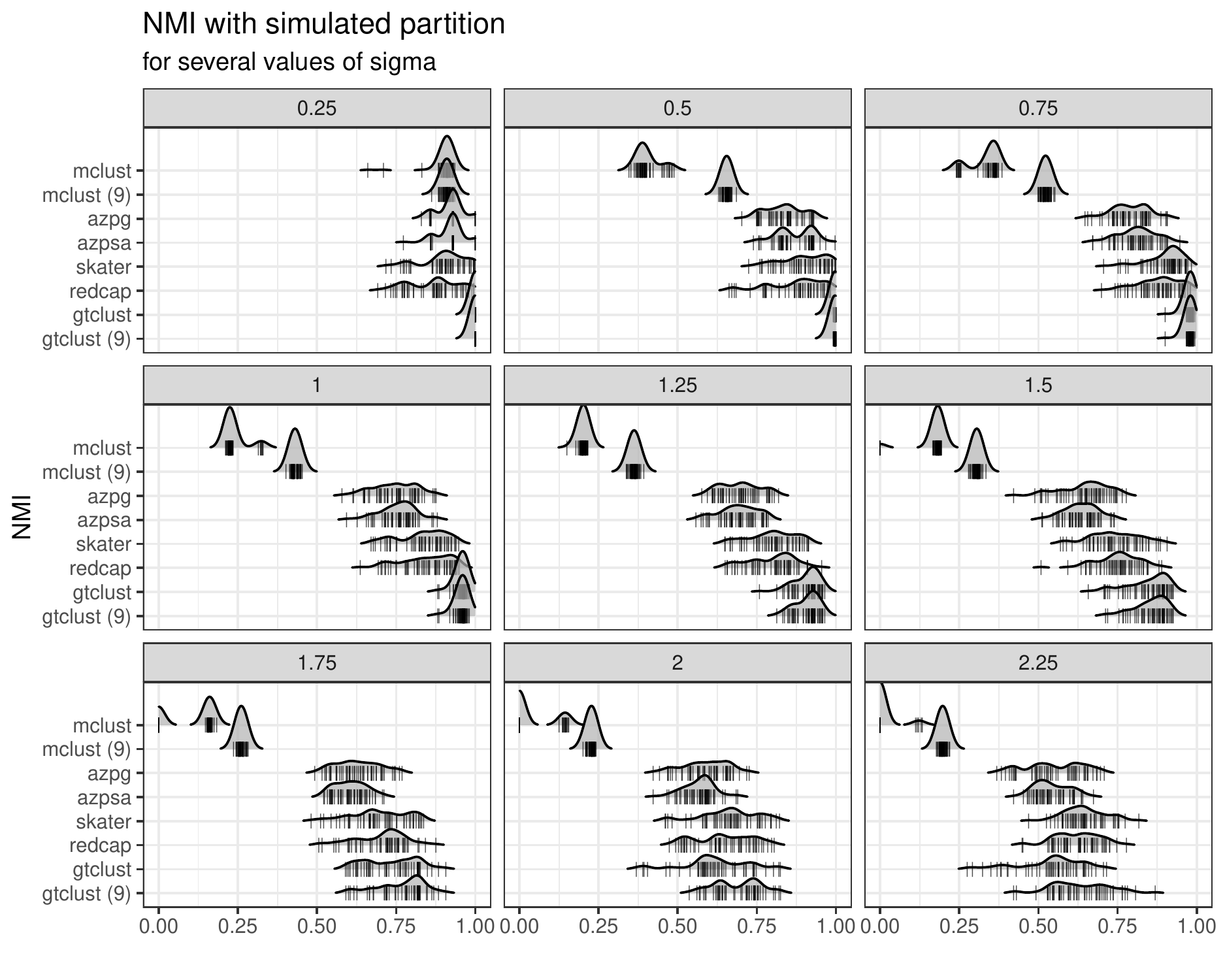}
\caption{\label{fig:resgridall}\footnotesize{Normalized Mutual Information with the simulated partition distribution for each algorithm and $\sigma$ value. Individual results are depicted by black lines.}}
\end{center}
\end{figure}

For each $\sigma$ value on the grid, 50 simulated records were generated and the results of each algorithm were recorded in terms of \textit{Normalised Mutual Information} (NMI) between the extracted partition and the simulated one. All algorithms were given the same contiguity graph (constructed with rook adjacency). The details of the results obtained are shown in Figure~\ref{fig:exresgrid}. The average performance of each algorithm is also given in Table~\ref{tab:simnmi}. The results obtained by \textbf{gtclust} are quite good, outperforming all the other methods until $\sigma$ reaches the value of 1.75, after this value the problems are quite hard with a very low signal to noise ratio and all the contiguity constrained methods reach similar performances. But below this value, \textbf{gtclust}, even without fixing the number of clusters to the true value, has better performances than the other methods (which need to know the number of clusters in advance) and is also more stable than the other approaches, as we can see in Figure~\ref{fig:resgridall}. Furthermore, in this range of settings, the number of clusters automatically found by \textbf{gtclust} is almost always equal to 9 or 8, as we can see in Table~\ref{tab:simK}. 

\begin{table}[ht!]
\label{tab:simnmi}
\begin{center}
\begin{tabular}{lrrrrrrrrr}
\hline\hline
$\sigma$ & 0.25 & 0.5 & 0.75 & 1 & 1.25 & 1.5 & 1.75 & 2 & 2.25\\
\hline
azpg & 0.92 & 0.83 & 0.79 & 0.74 & 0.70 & 0.64 & 0.63 & 0.59 & 0.55\\

azpsa & 0.92 & 0.87 & 0.81 & 0.75 & 0.69 & 0.63 & 0.61 & 0.57 & 0.54\\

gtclust & \textbf{1.00} & \textbf{1.00} & \textbf{0.98} & \textbf{0.96} & \textbf{0.91} & \textbf{0.84} & 0.74 & 0.62 & 0.54\\

gtclust (9) & \textbf{1.00} & \textbf{1.00} & \textbf{0.98} & \textbf{0.96} & \textbf{0.91} & \textbf{0.85} & \textbf{0.77} & 0.69 & 0.63\\

mclust & 0.90 & 0.40 & 0.33 & 0.24 & 0.20 & 0.17 & 0.10 & 0.04 & 0.01\\

mclust (9) & 0.91 & 0.65 & 0.52 & 0.43 & 0.36 & 0.31 & 0.26 & 0.23 & 0.20\\

redcap & 0.86 & 0.87 & 0.88 & 0.84 & 0.80 & 0.75 & 0.70 & 0.64 & 0.62\\

skater & 0.90 & 0.90 & 0.90 & 0.83 & 0.80 & 0.74 & 0.69 & 0.66 & 0.65\\
\hline\hline
\end{tabular}
\caption{\label{tab:sminmi}\footnotesize{Normalized Mutual Information with the simulated partition average over 50 simulations for each algorithm and $\sigma$ value. Results where gtclust outperforms significantly (according to a paired Wilcoxon tests) all the other methods are in bold.}}
\end{center}
\end{table}

\begin{table}[ht!]
\label{tab:simK}
\begin{center}
\begin{tabular}{lrrrrrrrrr}
\hline
$\sigma$ & 0.25 & 0.5 & 0.75 & 1.00 & 1.25 & 1.50 & 1.75 & 2.00 & 2.25 \\
\hline\hline
$p(\hat{K}=9) \% $ & 100 & 100 & 96 & 94 & 74 & 58 & 20 & 0 & 0 \\
average $\hat{K}$ & 9 & 9 & 9 & 8.98 & 8.80 & 8.42 & 7.14 & 5.82 & 5.02 \\
\hline\hline
\end{tabular}
\caption{\footnotesize{
Observed frequency of $\hat{K}=9$ for \textbf{gtclust} and average value of $\hat{K}$ over 50 simulations, with respect to $\sigma$. 
}}
\end{center}
\end{table}
With respect to the other approaches, the performance of the mixture models without contiguity constraints deteriorates rapidly. The solutions found by SKATER and REDCAP are comparable and slightly better than those found by the two AZP variants. These simulations thus highlight the interest of the proposed solution and of contiguity constraints.

\subsection{Regionalization of french mobility statistics}

The first results on real data that we describe come from the analysis of the mobility statistics of the 2018 French census, provided by INSEE\footnote{see https://www.insee.fr/fr/statistiques/5650708}. Among the information collected for the census, several pieces of information on the main mode of transport are collected, and we analysed the share of residents using: car/transit/motorised two-wheelers/bicycle/footwalking or any other mode as the main mode of transport in the " region centre " of France at the IRIS level. IRIS are geographical units that divide the French municipalities into regions of around 2000 inhabitants. There are 2150 IRIS in the dataset used and they are described by the six variables mentioned above. Geographical units with less than 50 inhabitants were smoothed with a weighted average of their neighbourhood values.  We considered the data as compositional and transformed them using an additive log-ratio transformation with the car variable as the reference variable, leaving 5 variables to be analysed. A multivariate Gaussian mixture model with diagonal covariance matrices and normal gamma prior was then applied to these transformed data. The prior parameters were set to $(\tau = 0.01$, $\kappa = 1$, $\beta = 0.64$, $\mu = \bar{\mathbf{x}}$). 

\begin{figure}[ht!]
\begin{center}
\includegraphics[width=0.45\textwidth]{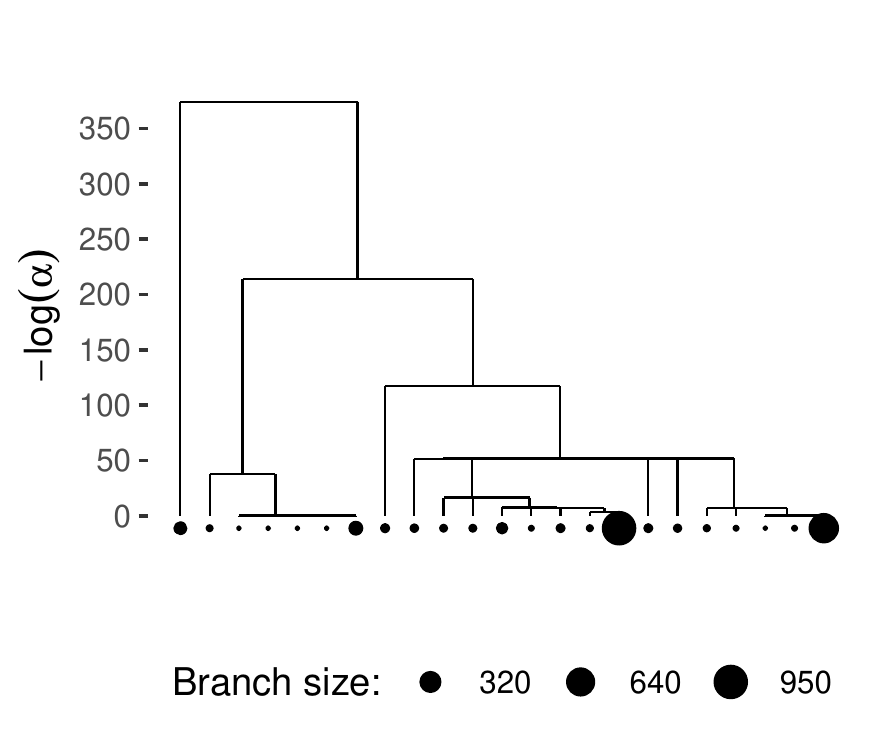}
\caption{\label{fig:treemob}\footnotesize{Bayesian dendrogram on the "region centre mobility statics dataset".}}
\end{center}
\end{figure}

\begin{figure}[ht!]
\begin{center}
\includegraphics[width=0.95\textwidth]{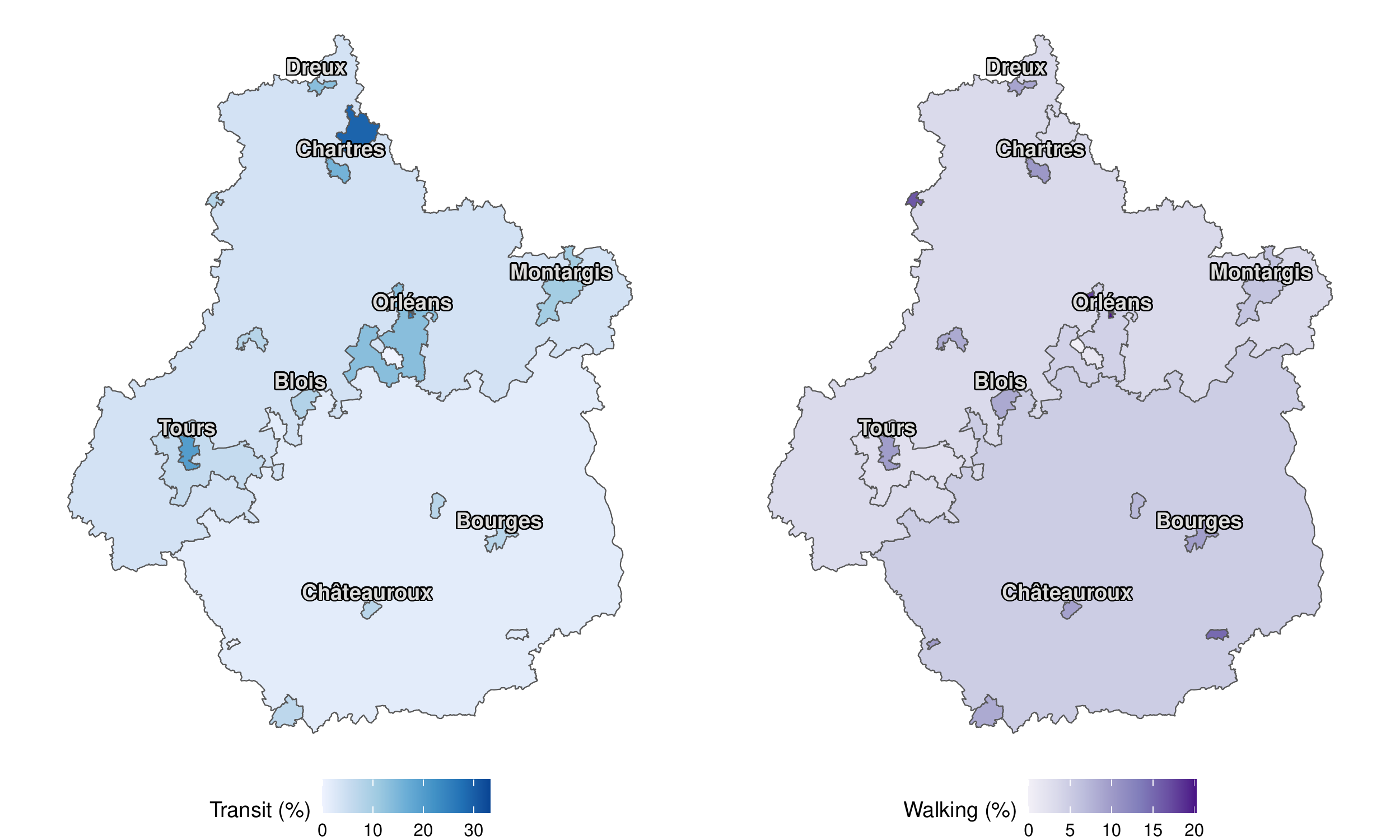}
\caption{\label{fig:clustmob}\footnotesize{Regionalization results on the "region centre mobility statics dataset".}}
\end{center}
\end{figure}

The dendrogram of the solution found is shown in Figure~\ref{fig:treemob} and starts with 23 clusters. As shown in Figure~\ref{fig:clustmob}, the main characteristics of the territory are clearly extracted by the algorithm, the main agglomerations (highlighted with labels) form specific clusters, the two largest (Tours and Orléans) are even represented by two clusters (one for the inner city and another for the suburbs), which is easily explained by the higher share of walking and transit use in these regions. Two large clusters divide the region in a north-south direction, with a lower use of transit (and a higher use of walking) in the southern part of the region. Finally, a final cluster in the north-east of the map presents a heavy use of transit and can be explained by important commuting flows with Paris and made to a large extent by transit.

\subsection{Traffic speed clustering}

The second application on real data that we have carried out deals with traffic speeds on a road network. Such an application is of interest because finding homogeneous traffic zones in road networks can be very helpful for traffic control. An important open question in traffic theory for the application of the Macroscopic Fundamental Diagram (MFD), a classical tool in traffic control, is how best to segment an urban network into regional subnetworks and how to treat endogenous heterogeneity in the spatial distribution of congestion \citep{Ji2012,Haghbayan21}. This segmentation must fulfil several properties: first, the regions must be of reasonable and similar size, and second, they must be topologically connected and compact. Finally, the traffic conditions in each region must be approximately homogeneous (i.e. congestion must be approximately homogeneous in the region). Contiguity constrained clustering therefore seems a natural way to solve this problem and we have investigated the use of the proposed algorithm in this context.

\begin{figure}[ht!]
\begin{center}
\includegraphics[width=0.95\textwidth]{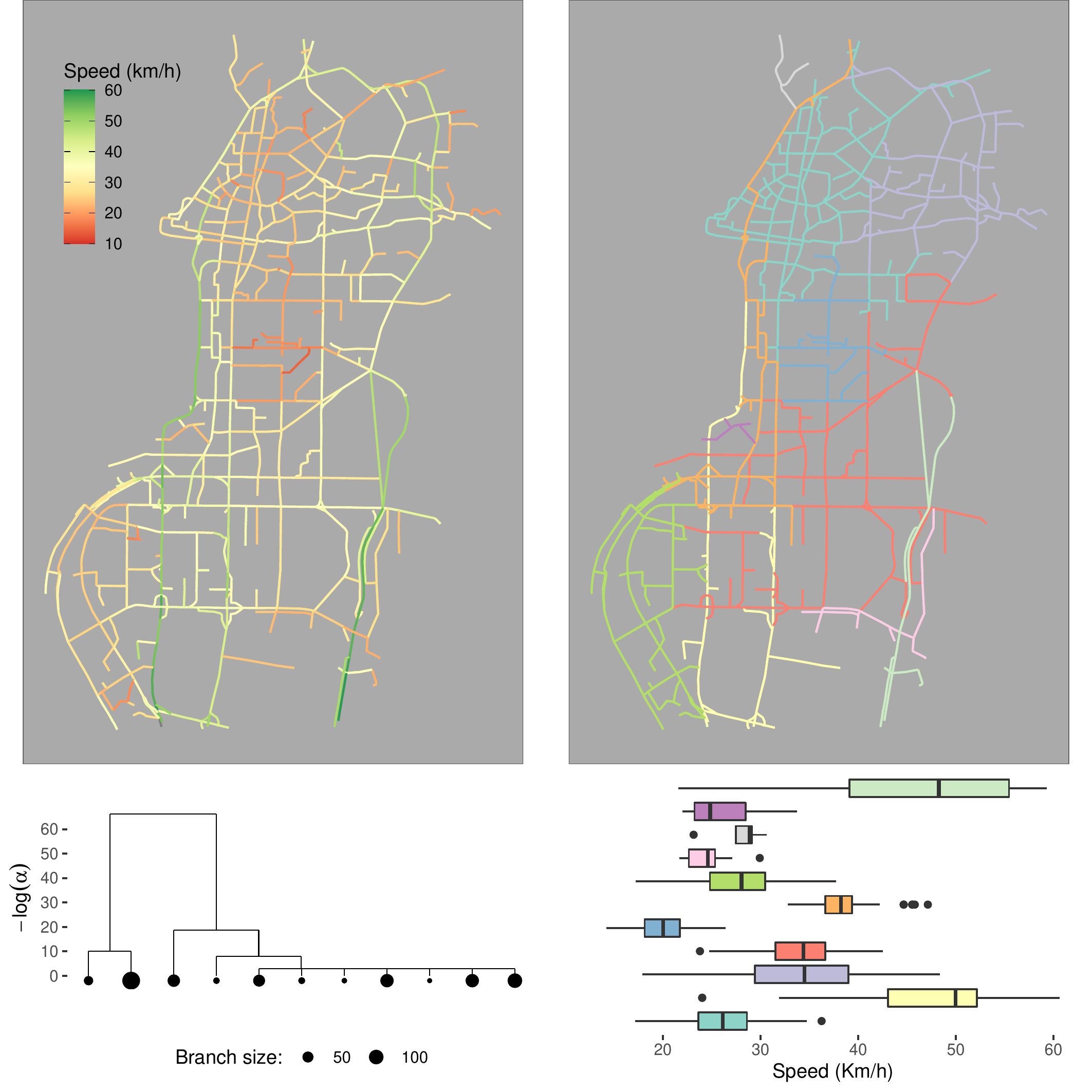}
\caption{\label{fig:shenzhenclust}\footnotesize{Network clustering results on the "Shenzhen speeds dataset". Data from 9th September 2011, average speed during the 7h30-7h45 timeslot; speed per link map (top-left) ; clusters map (top-right) ; dendrogram (bottom-left) ; boxplots of speeds distributions per clusters (bottom-right).}}
\end{center}
\end{figure}

The data used in this section is one of the benchmark datasets used to study this task \citep{Bellocchi2019}. This dataset concerns the road network of Shenzhen, where the average speed of road segments is estimated every 5 minutes from map-matched traces of about 20,000 taxi GPS points collected over three days in 2011. In order to replicate the preprocessing steps used in previous studies \citep{Haghbayan21}, the following preprocessing steps were performed: nodes with no structural role (i.e. that are not intersections) were smoothed (their incoming edges were merged and the associated speed value was taken as the mean of the speeds of the combined edges); average speed values were also computed over 15-minute time intervals. Finally, the graph between road segments was constructed by connecting any two segments connected by an intersection, as shown in Figure~\ref{fig:nets}. A Gaussian mixture model with Norma-Gamma conjugate prior $\mu_k\sim\mathcal{N}\left(\mu,(\tau. V_k)^{-1}\right),V_k\sim Gam(\kappa,\beta),$ with prior parameters ($\tau = 0.01$, $\kappa = 1$, $\beta = 0.1s^2 \approx 0.62$, $\mu = \bar{x} \approx 9$) and the velocities in $m/s$ were then used to cluster the data set.

We present in Figure~\ref{fig:shenzhenclust} the clustering results obtained over the 7h30-7h45 time slot of the 9th of September. The figure shows the observed velocity in a first map and the extracted clusters in another. These two maps are accompanied by the computed dendrogram and per cluster box plots of the observed velocities. The uniform prior leads to 11 clusters, from the dendrogram we can see that the next interesting solutions contain only 5 clusters. The traffic conditions are homogeneous in each cluster, two clusters correspond to free flow zones with an average speed around 50 km/h, the others correspond to congested zones with speeds around 20 km/h or 35 km/h.  When aggregated to the next level of interest in the dendrogram, the remaining clusters are the pink, light green and dark blue clusters, the others being merged into a single "non-congested" cluster. Again, the segmentation obtained by clustering seems to be quite coherent and meaningful, showing the interest of the proposal for this type of applications.

\section{Conclusion and future works}
\label{sec:conclu}

This paper has introduced a Bayesian approach to contiguity constrained clustering that allows semi-automatic tuning of the model complexity (number of clusters), together with an efficient algorithm for finding a MAP partition and building a Bayesian dendrogram from it. A simulation study has shown the interest of this algorithm compared to other solutions for contiguity constrained clustering and its ability to recover the number of clusters. Finally, two experiments on real data have demonstrated the applicability of the proposal on real test cases.  From an algorithmic point of view, it would be interesting to compare the hierarchical approach used in this proposal with other solutions based, for example, on local swap moves (\textit{ie.} in the sense of the Louvain or Leiden algorithms), and to compare the advantages or disadvantages of these two solutions on benchmark datasets. From a modelling point of view, we have mainly illustrated the proposal with Gaussian Mixture Models, but as the proposal is quite general, other models should be investigated (Poisson, Mulitnomial, ...). In this spirit, we are currently working on a solution based on the Laplace approximation to deal with more complex observation models where conjugate priors do not exist.

\bibliographystyle{johd}
\bibliography{references.bib}

\begin{thebibliography}{}

\bibitem [\protect \citeauthoryear {%
Angriman%
, Predari%
, van~der Grinten%
\BCBL {}\ \BBA {} Meyerhenke%
}{%
Angriman%
\ \protect \BOthers {.}}{%
{\protect \APACyear {2020}}%
}]{%
Angriman20}
\APACinsertmetastar {%
Angriman20}%
\begin{APACrefauthors}%
Angriman, E.%
, Predari, M.%
, van~der Grinten, A.%
\BCBL {}\ \BBA {} Meyerhenke, H.%
\end{APACrefauthors}%
\unskip\
\newblock
\APACrefYearMonthDay{2020}{}{}.
\newblock
\APACrefbtitle {Approximation of the Diagonal of a Laplacian's Pseudoinverse
  for Complex Network Analysis.} {Approximation of the diagonal of a
  laplacian's pseudoinverse for complex network analysis.}
\newblock
\APACaddressPublisher{}{arXiv}.
\newblock
\begin{APACrefURL} \url{https://arxiv.org/abs/2006.13679} \end{APACrefURL}
\newblock
\doi{10.48550/ARXIV.2006.13679}
\PrintBackRefs{\CurrentBib}

\bibitem [\protect \citeauthoryear {%
Anselin%
}{%
Anselin%
}{%
{\protect \APACyear {2001}}%
}]{%
Anselin01}
\APACinsertmetastar {%
Anselin01}%
\begin{APACrefauthors}%
Anselin, L.%
\end{APACrefauthors}%
\unskip\
\newblock
\APACrefYearMonthDay{2001}{}{}.
\newblock
{\BBOQ}\APACrefatitle {Spatial econometrics} {Spatial econometrics}.{\BBCQ}
\newblock
\BIn{} \APACrefbtitle {A companion to theoretical econometrics} {A companion to
  theoretical econometrics}\ (\BVOL\ 310330, \BPGS\ 310--330).
\PrintBackRefs{\CurrentBib}

\bibitem [\protect \citeauthoryear {%
Assunção%
, Neves%
, Câmara%
\BCBL {}\ \BBA {} Freitas%
}{%
Assunção%
\ \protect \BOthers {.}}{%
{\protect \APACyear {2006}}%
}]{%
Assuncao2006}
\APACinsertmetastar {%
Assuncao2006}%
\begin{APACrefauthors}%
Assunção, R\BPBI M.%
, Neves, M\BPBI C.%
, Câmara, G.%
\BCBL {}\ \BBA {} Freitas, C\BPBI D\BPBI C.%
\end{APACrefauthors}%
\unskip\
\newblock
\APACrefYearMonthDay{2006}{}{}.
\newblock
{\BBOQ}\APACrefatitle {Efficient regionalization techniques for
  socio‐economic geographical units using minimum spanning trees} {Efficient
  regionalization techniques for socio‐economic geographical units using
  minimum spanning trees}.{\BBCQ}
\newblock
\APACjournalVolNumPages{International Journal of Geographical Information
  Science}{20}{7}{797-811}.
\newblock
\begin{APACrefURL} \url{https://doi.org/10.1080/13658810600665111}
  \end{APACrefURL}
\newblock
\doi{10.1080/13658810600665111}
\PrintBackRefs{\CurrentBib}

\bibitem [\protect \citeauthoryear {%
Barry%
\ \BBA {} Hartigan%
}{%
Barry%
\ \BBA {} Hartigan%
}{%
{\protect \APACyear {1993}}%
}]{%
Barry93}
\APACinsertmetastar {%
Barry93}%
\begin{APACrefauthors}%
Barry, D.%
\BCBT {}\ \BBA {} Hartigan, J\BPBI A.%
\end{APACrefauthors}%
\unskip\
\newblock
\APACrefYearMonthDay{1993}{}{}.
\newblock
{\BBOQ}\APACrefatitle {A Bayesian Analysis for Change Point Problems} {A
  bayesian analysis for change point problems}.{\BBCQ}
\newblock
\APACjournalVolNumPages{Journal of the American Statistical
  Association}{88}{421}{309-319}.
\newblock
\begin{APACrefURL} \url{https://doi.org/10.1080/01621459.1993.10594323}
  \end{APACrefURL}
\newblock
\doi{10.1080/01621459.1993.10594323}
\PrintBackRefs{\CurrentBib}

\bibitem [\protect \citeauthoryear {%
Bates%
\ \BBA {} Maechler%
}{%
Bates%
\ \BBA {} Maechler%
}{%
{\protect \APACyear {2019}}%
}]{%
Bates2019}
\APACinsertmetastar {%
Bates2019}%
\begin{APACrefauthors}%
Bates, D.%
\BCBT {}\ \BBA {} Maechler, M.%
\end{APACrefauthors}%
\unskip\
\newblock
\APACrefYearMonthDay{2019}{}{}.
\newblock
{\BBOQ}\APACrefatitle {Matrix: Sparse and Dense Matrix Classes and Methods}
  {Matrix: Sparse and dense matrix classes and methods}{\BBCQ}\
  [\bibcomputersoftwaremanual].
\newblock
\begin{APACrefURL} \url{https://CRAN.R-project.org/package=Matrix}
  \end{APACrefURL}
\newblock
\APACrefnote{R package version 1.2-17}
\PrintBackRefs{\CurrentBib}

\bibitem [\protect \citeauthoryear {%
Bellocchi%
\ \BBA {} Geroliminis%
}{%
Bellocchi%
\ \BBA {} Geroliminis%
}{%
{\protect \APACyear {2019}}%
}]{%
Bellocchi2019}
\APACinsertmetastar {%
Bellocchi2019}%
\begin{APACrefauthors}%
Bellocchi, L.%
\BCBT {}\ \BBA {} Geroliminis, N.%
\end{APACrefauthors}%
\unskip\
\newblock
\APACrefYearMonthDay{2019}{3}{}.
\newblock
\APACrefbtitle {Shenzhen whole day Speeds.} {Shenzhen whole day speeds.}
\newblock
\begin{APACrefURL}
  \url{https://figshare.com/articles/dataset/Shenzhen_whole_day_Speeds/7212230}
  \end{APACrefURL}
\newblock
\doi{10.6084/m9.figshare.7212230.v2}
\PrintBackRefs{\CurrentBib}

\bibitem [\protect \citeauthoryear {%
Biernacki%
, Celeux%
\BCBL {}\ \BBA {} Govaert%
}{%
Biernacki%
\ \protect \BOthers {.}}{%
{\protect \APACyear {2000}}%
}]{%
Biernacki2000}
\APACinsertmetastar {%
Biernacki2000}%
\begin{APACrefauthors}%
Biernacki, C.%
, Celeux, G.%
\BCBL {}\ \BBA {} Govaert, G.%
\end{APACrefauthors}%
\unskip\
\newblock
\APACrefYearMonthDay{2000}{}{}.
\newblock
{\BBOQ}\APACrefatitle {Assessing a mixture model for clustering with the
  integrated completed likelihood} {Assessing a mixture model for clustering
  with the integrated completed likelihood}.{\BBCQ}
\newblock
\APACjournalVolNumPages{IEEE Transaction on Pattern Analysis and Machine
  Intelligence}{7}{}{719-725}.
\PrintBackRefs{\CurrentBib}

\bibitem [\protect \citeauthoryear {%
Biernacki%
, Celeux%
\BCBL {}\ \BBA {} Govaert%
}{%
Biernacki%
\ \protect \BOthers {.}}{%
{\protect \APACyear {2010}}%
}]{%
Biernacki2010}
\APACinsertmetastar {%
Biernacki2010}%
\begin{APACrefauthors}%
Biernacki, C.%
, Celeux, G.%
\BCBL {}\ \BBA {} Govaert, G.%
\end{APACrefauthors}%
\unskip\
\newblock
\APACrefYearMonthDay{2010}{}{}.
\newblock
{\BBOQ}\APACrefatitle {Exact and monte carlo calculations of integrated
  likelihoods for the latent class model} {Exact and monte carlo calculations
  of integrated likelihoods for the latent class model}.{\BBCQ}
\newblock
\APACjournalVolNumPages{Journal of Statistical Planning and
  Inference}{140}{}{2991-3002}.
\PrintBackRefs{\CurrentBib}

\bibitem [\protect \citeauthoryear {%
Blondel%
, Guillaume%
, Lambiotte%
\BCBL {}\ \BBA {} Lefebvre%
}{%
Blondel%
\ \protect \BOthers {.}}{%
{\protect \APACyear {2008}}%
}]{%
Blondel08}
\APACinsertmetastar {%
Blondel08}%
\begin{APACrefauthors}%
Blondel, V\BPBI D.%
, Guillaume, J\BHBI L.%
, Lambiotte, R.%
\BCBL {}\ \BBA {} Lefebvre, E.%
\end{APACrefauthors}%
\unskip\
\newblock
\APACrefYearMonthDay{2008}{}{}.
\newblock
{\BBOQ}\APACrefatitle {Fast unfolding of communities in large networks} {Fast
  unfolding of communities in large networks}.{\BBCQ}
\newblock
\APACjournalVolNumPages{Journal of statistical mechanics: theory and
  experiment}{2008}{10}{P10008}.
\PrintBackRefs{\CurrentBib}

\bibitem [\protect \citeauthoryear {%
Cayley%
}{%
Cayley%
}{%
{\protect \APACyear {1889}}%
}]{%
Cayley1889}
\APACinsertmetastar {%
Cayley1889}%
\begin{APACrefauthors}%
Cayley, A.%
\end{APACrefauthors}%
\unskip\
\newblock
\APACrefYearMonthDay{1889}{}{}.
\newblock
{\BBOQ}\APACrefatitle {A theorem on trees} {A theorem on trees}.{\BBCQ}
\newblock
\APACjournalVolNumPages{Quaterly Journal of Mathematics}{23}{}{376--378}.
\PrintBackRefs{\CurrentBib}

\bibitem [\protect \citeauthoryear {%
Chavent%
, Kuentz-Simonet%
, Labenne%
\BCBL {}\ \BBA {} Saracco%
}{%
Chavent%
\ \protect \BOthers {.}}{%
{\protect \APACyear {2018}}%
}]{%
Chavent2018}
\APACinsertmetastar {%
Chavent2018}%
\begin{APACrefauthors}%
Chavent, M.%
, Kuentz-Simonet, V.%
, Labenne, A.%
\BCBL {}\ \BBA {} Saracco, J.%
\end{APACrefauthors}%
\unskip\
\newblock
\APACrefYearMonthDay{2018}{dec}{}.
\newblock
{\BBOQ}\APACrefatitle {ClustGeo: An R Package for Hierarchical Clustering with
  Spatial Constraints} {Clustgeo: An r package for hierarchical clustering with
  spatial constraints}.{\BBCQ}
\newblock
\APACjournalVolNumPages{Comput. Stat.}{33}{4}{1799–1822}.
\newblock
\begin{APACrefURL} \url{https://doi.org/10.1007/s00180-018-0791-1}
  \end{APACrefURL}
\newblock
\doi{10.1007/s00180-018-0791-1}
\PrintBackRefs{\CurrentBib}

\bibitem [\protect \citeauthoryear {%
Chen%
, Davis%
, Hager%
\BCBL {}\ \BBA {} Rajamanickam%
}{%
Chen%
\ \protect \BOthers {.}}{%
{\protect \APACyear {2008}}%
}]{%
Chen08}
\APACinsertmetastar {%
Chen08}%
\begin{APACrefauthors}%
Chen, Y.%
, Davis, T.%
, Hager, W.%
\BCBL {}\ \BBA {} Rajamanickam, S.%
\end{APACrefauthors}%
\unskip\
\newblock
\APACrefYearMonthDay{2008}{}{}.
\newblock
{\BBOQ}\APACrefatitle {Algorithm 887: CHOLMOD, supernodal sparse Cholesky
  factorization and update/downdate} {Algorithm 887: Cholmod, supernodal sparse
  cholesky factorization and update/downdate}.{\BBCQ}
\newblock
\APACjournalVolNumPages{ACM Trans. on Mathematical
  Software}{35}{}{22:1--22:14}.
\newblock
\doi{10.1145/1391989.1391995}
\PrintBackRefs{\CurrentBib}

\bibitem [\protect \citeauthoryear {%
Christophe%
, Alia%
, Pierre%
, Guillem%
\BCBL {}\ \BBA {} Nathalie%
}{%
Christophe%
\ \protect \BOthers {.}}{%
{\protect \APACyear {2019}}%
}]{%
Ambroise2019}
\APACinsertmetastar {%
Ambroise2019}%
\begin{APACrefauthors}%
Christophe, A.%
, Alia, D.%
, Pierre, N.%
, Guillem, R.%
\BCBL {}\ \BBA {} Nathalie, V.%
\end{APACrefauthors}%
\unskip\
\newblock
\APACrefYearMonthDay{2019}{}{}.
\newblock
{\BBOQ}\APACrefatitle {Adjacency-constrained hierarchical clustering of a band
  similarity matrix with application to genomics} {Adjacency-constrained
  hierarchical clustering of a band similarity matrix with application to
  genomics}.{\BBCQ}
\newblock
\APACjournalVolNumPages{Algorithms for Molecular Biology}{14}{1}{22}.
\PrintBackRefs{\CurrentBib}

\bibitem [\protect \citeauthoryear {%
C{\^o}me%
, Latouche%
, Jouvin%
\BCBL {}\ \BBA {} Bouveyron%
}{%
C{\^o}me%
\ \protect \BOthers {.}}{%
{\protect \APACyear {2021}}%
}]{%
Come21}
\APACinsertmetastar {%
Come21}%
\begin{APACrefauthors}%
C{\^o}me, E.%
, Latouche, P.%
, Jouvin, N.%
\BCBL {}\ \BBA {} Bouveyron, C.%
\end{APACrefauthors}%
\unskip\
\newblock
\APACrefYearMonthDay{2021}{}{}.
\newblock
{\BBOQ}\APACrefatitle {{Hierarchical clustering with discrete latent variable
  models and the integrated classification likelihood}} {{Hierarchical
  clustering with discrete latent variable models and the integrated
  classification likelihood}}.{\BBCQ}
\newblock
\APACjournalVolNumPages{{Advances in Data Analysis and Classification}}{}{}{}.
\newblock
\begin{APACrefURL} \url{https://hal.archives-ouvertes.fr/hal-02530705}
  \end{APACrefURL}
\newblock
\doi{10.1007/s11634-021-00440-z}
\PrintBackRefs{\CurrentBib}

\bibitem [\protect \citeauthoryear {%
Davis%
\ \BBA {} Hager%
}{%
Davis%
\ \BBA {} Hager%
}{%
{\protect \APACyear {1999}}%
}]{%
Davis99}
\APACinsertmetastar {%
Davis99}%
\begin{APACrefauthors}%
Davis, T\BPBI A.%
\BCBT {}\ \BBA {} Hager, W\BPBI W.%
\end{APACrefauthors}%
\unskip\
\newblock
\APACrefYearMonthDay{1999}{}{}.
\newblock
{\BBOQ}\APACrefatitle {Modifying a Sparse Cholesky Factorization} {Modifying a
  sparse cholesky factorization}.{\BBCQ}
\newblock
\APACjournalVolNumPages{SIAM Journal on Matrix Analysis and
  Applications}{20}{}{606--627}.
\newblock
\doi{10.1137/S0895479897321076}
\PrintBackRefs{\CurrentBib}

\bibitem [\protect \citeauthoryear {%
Davis%
\ \BBA {} Hager%
}{%
Davis%
\ \BBA {} Hager%
}{%
{\protect \APACyear {2001}}%
}]{%
Davis01}
\APACinsertmetastar {%
Davis01}%
\begin{APACrefauthors}%
Davis, T\BPBI A.%
\BCBT {}\ \BBA {} Hager, W\BPBI W.%
\end{APACrefauthors}%
\unskip\
\newblock
\APACrefYearMonthDay{2001}{}{}.
\newblock
{\BBOQ}\APACrefatitle {Multiple-Rank Modifications of a Sparse Cholesky
  Factorization} {Multiple-rank modifications of a sparse cholesky
  factorization}.{\BBCQ}
\newblock
\APACjournalVolNumPages{SIAM Journal on Matrix Analysis and
  Applications}{22}{}{997--1013}.
\newblock
\doi{10.1137/S0895479899357346}
\PrintBackRefs{\CurrentBib}

\bibitem [\protect \citeauthoryear {%
Davis%
\ \BBA {} Hager%
}{%
Davis%
\ \BBA {} Hager%
}{%
{\protect \APACyear {2005}}%
}]{%
Davis05}
\APACinsertmetastar {%
Davis05}%
\begin{APACrefauthors}%
Davis, T\BPBI A.%
\BCBT {}\ \BBA {} Hager, W\BPBI W.%
\end{APACrefauthors}%
\unskip\
\newblock
\APACrefYearMonthDay{2005}{}{}.
\newblock
{\BBOQ}\APACrefatitle {Row Modifications of a Sparse Cholesky Factorization
  SIAM Journal on Matrix Analysis and Applications} {Row modifications of a
  sparse cholesky factorization siam journal on matrix analysis and
  applications}.{\BBCQ}
\newblock
\APACjournalVolNumPages{SIAM Journal on Matrix Analysis and
  Applications}{26}{}{621--639}.
\newblock
\doi{10.1137/S089547980343641X}
\PrintBackRefs{\CurrentBib}

\bibitem [\protect \citeauthoryear {%
Diaconis%
\ \BBA {} Ylvisaker%
}{%
Diaconis%
\ \BBA {} Ylvisaker%
}{%
{\protect \APACyear {1979}}%
}]{%
diaconis1979conjugate}
\APACinsertmetastar {%
diaconis1979conjugate}%
\begin{APACrefauthors}%
Diaconis, P.%
\BCBT {}\ \BBA {} Ylvisaker, D.%
\end{APACrefauthors}%
\unskip\
\newblock
\APACrefYearMonthDay{1979}{}{}.
\newblock
{\BBOQ}\APACrefatitle {Conjugate priors for exponential families} {Conjugate
  priors for exponential families}.{\BBCQ}
\newblock
\APACjournalVolNumPages{The Annals of statistics}{}{}{269--281}.
\PrintBackRefs{\CurrentBib}

\bibitem [\protect \citeauthoryear {%
Eddelbuettel%
\ \BBA {} Balamuta%
}{%
Eddelbuettel%
\ \BBA {} Balamuta%
}{%
{\protect \APACyear {2017}}%
}]{%
Eddelbuettel2017}
\APACinsertmetastar {%
Eddelbuettel2017}%
\begin{APACrefauthors}%
Eddelbuettel, D.%
\BCBT {}\ \BBA {} Balamuta, J\BPBI J.%
\end{APACrefauthors}%
\unskip\
\newblock
\APACrefYearMonthDay{2017}{aug}{}.
\newblock
{\BBOQ}\APACrefatitle {{Extending extit{R} with extit{C++}: A Brief
  Introduction to extit{Rcpp}}} {{Extending extit{R} with extit{C++}: A Brief
  Introduction to extit{Rcpp}}}.{\BBCQ}
\newblock
\APACjournalVolNumPages{PeerJ Preprints}{5}{}{e3188v1}.
\newblock
\begin{APACrefURL} \url{https://doi.org/10.7287/peerj.preprints.3188v1}
  \end{APACrefURL}
\newblock
\doi{10.7287/peerj.preprints.3188v1}
\PrintBackRefs{\CurrentBib}

\bibitem [\protect \citeauthoryear {%
Eddelbuettel%
\ \BBA {} Sanderson%
}{%
Eddelbuettel%
\ \BBA {} Sanderson%
}{%
{\protect \APACyear {2014}}%
}]{%
Eddelbuettel2014}
\APACinsertmetastar {%
Eddelbuettel2014}%
\begin{APACrefauthors}%
Eddelbuettel, D.%
\BCBT {}\ \BBA {} Sanderson, C.%
\end{APACrefauthors}%
\unskip\
\newblock
\APACrefYearMonthDay{2014}{March}{}.
\newblock
{\BBOQ}\APACrefatitle {RcppArmadillo: Accelerating R with high-performance C++
  linear algebra} {Rcpparmadillo: Accelerating r with high-performance c++
  linear algebra}.{\BBCQ}
\newblock
\APACjournalVolNumPages{Computational Statistics and Data
  Analysis}{71}{}{1054--1063}.
\newblock
\begin{APACrefURL} \url{http://dx.doi.org/10.1016/j.csda.2013.02.005}
  \end{APACrefURL}
\PrintBackRefs{\CurrentBib}

\bibitem [\protect \citeauthoryear {%
Gordon%
}{%
Gordon%
}{%
{\protect \APACyear {1996}}%
}]{%
Gordon96}
\APACinsertmetastar {%
Gordon96}%
\begin{APACrefauthors}%
Gordon, A.%
\end{APACrefauthors}%
\unskip\
\newblock
\APACrefYearMonthDay{1996}{}{}.
\newblock
{\BBOQ}\APACrefatitle {A survey of constrained classification} {A survey of
  constrained classification}.{\BBCQ}
\newblock
\APACjournalVolNumPages{Computational Statistics \& Data
  Analysis}{21}{}{17–29}.
\PrintBackRefs{\CurrentBib}

\bibitem [\protect \citeauthoryear {%
Grimm%
}{%
Grimm%
}{%
{\protect \APACyear {1987}}%
}]{%
Grim87}
\APACinsertmetastar {%
Grim87}%
\begin{APACrefauthors}%
Grimm, E\BPBI C.%
\end{APACrefauthors}%
\unskip\
\newblock
\APACrefYearMonthDay{1987}{}{}.
\newblock
{\BBOQ}\APACrefatitle {CONISS: a FORTRAN 77 program for stratigraphically
  constrained analysis by the method of incremental sum of squares} {Coniss: a
  fortran 77 program for stratigraphically constrained analysis by the method
  of incremental sum of squares}.{\BBCQ}
\newblock
\APACjournalVolNumPages{Computers \& Geosciences}{13}{}{13–35}.
\PrintBackRefs{\CurrentBib}

\bibitem [\protect \citeauthoryear {%
Guo%
}{%
Guo%
}{%
{\protect \APACyear {2008}}%
}]{%
Guo2008}
\APACinsertmetastar {%
Guo2008}%
\begin{APACrefauthors}%
Guo, D.%
\end{APACrefauthors}%
\unskip\
\newblock
\APACrefYearMonthDay{2008}{}{}.
\newblock
{\BBOQ}\APACrefatitle {Regionalization with dynamically constrained
  agglomerative clustering and partitioning (REDCAP)} {Regionalization with
  dynamically constrained agglomerative clustering and partitioning
  (redcap)}.{\BBCQ}
\newblock
\APACjournalVolNumPages{International Journal of Geographical Information
  Science}{22}{7}{801-823}.
\newblock
\begin{APACrefURL} \url{https://doi.org/10.1080/13658810701674970}
  \end{APACrefURL}
\newblock
\doi{10.1080/13658810701674970}
\PrintBackRefs{\CurrentBib}

\bibitem [\protect \citeauthoryear {%
Haghbayan%
, Geroliminis%
\BCBL {}\ \BBA {} Akbarzadeh%
}{%
Haghbayan%
\ \protect \BOthers {.}}{%
{\protect \APACyear {2021}}%
}]{%
Haghbayan21}
\APACinsertmetastar {%
Haghbayan21}%
\begin{APACrefauthors}%
Haghbayan, S\BPBI A.%
, Geroliminis, N.%
\BCBL {}\ \BBA {} Akbarzadeh, M.%
\end{APACrefauthors}%
\unskip\
\newblock
\APACrefYearMonthDay{2021}{11}{}.
\newblock
{\BBOQ}\APACrefatitle {Community detection in large scale congested urban road
  networks} {Community detection in large scale congested urban road
  networks}.{\BBCQ}
\newblock
\APACjournalVolNumPages{PLOS ONE}{16}{11}{1-14}.
\newblock
\begin{APACrefURL} \url{https://doi.org/10.1371/journal.pone.0260201}
  \end{APACrefURL}
\newblock
\doi{10.1371/journal.pone.0260201}
\PrintBackRefs{\CurrentBib}

\bibitem [\protect \citeauthoryear {%
Hartigan%
}{%
Hartigan%
}{%
{\protect \APACyear {1990}}%
}]{%
Hartigan90}
\APACinsertmetastar {%
Hartigan90}%
\begin{APACrefauthors}%
Hartigan, J.%
\end{APACrefauthors}%
\unskip\
\newblock
\APACrefYearMonthDay{1990}{}{}.
\newblock
{\BBOQ}\APACrefatitle {Partition models} {Partition models}.{\BBCQ}
\newblock
\APACjournalVolNumPages{Communications in Statistics - Theory and
  Methods}{19}{8}{2745-2756}.
\newblock
\begin{APACrefURL} \url{https://doi.org/10.1080/03610929008830345}
  \end{APACrefURL}
\newblock
\doi{10.1080/03610929008830345}
\PrintBackRefs{\CurrentBib}

\bibitem [\protect \citeauthoryear {%
Hayashi%
, Akiba%
\BCBL {}\ \BBA {} Yoshida%
}{%
Hayashi%
\ \protect \BOthers {.}}{%
{\protect \APACyear {2016}}%
}]{%
Takanori16}
\APACinsertmetastar {%
Takanori16}%
\begin{APACrefauthors}%
Hayashi, T.%
, Akiba, T.%
\BCBL {}\ \BBA {} Yoshida, Y.%
\end{APACrefauthors}%
\unskip\
\newblock
\APACrefYearMonthDay{2016}{}{}.
\newblock
{\BBOQ}\APACrefatitle {Efficient Algorithms for Spanning Tree Centrality}
  {Efficient algorithms for spanning tree centrality}.{\BBCQ}
\newblock
\BIn{} \APACrefbtitle {Proceedings of the Twenty-Fifth International Joint
  Conference on Artificial Intelligence (IJCAI-16)} {Proceedings of the
  twenty-fifth international joint conference on artificial intelligence
  (ijcai-16)}\ (\BPGS\ 3733--3739).
\PrintBackRefs{\CurrentBib}

\bibitem [\protect \citeauthoryear {%
Hegarty%
\ \BBA {} Barry%
}{%
Hegarty%
\ \BBA {} Barry%
}{%
{\protect \APACyear {2008}}%
}]{%
Hegarty08}
\APACinsertmetastar {%
Hegarty08}%
\begin{APACrefauthors}%
Hegarty, A.%
\BCBT {}\ \BBA {} Barry, D.%
\end{APACrefauthors}%
\unskip\
\newblock
\APACrefYearMonthDay{2008}{}{}.
\newblock
{\BBOQ}\APACrefatitle {Bayesian disease mapping using product partition models}
  {Bayesian disease mapping using product partition models}.{\BBCQ}
\newblock
\APACjournalVolNumPages{Statistics in Medicine}{27}{19}{3868-3893}.
\newblock
\begin{APACrefURL}
  \url{https://onlinelibrary.wiley.com/doi/abs/10.1002/sim.3253}
  \end{APACrefURL}
\newblock
\doi{https://doi.org/10.1002/sim.3253}
\PrintBackRefs{\CurrentBib}

\bibitem [\protect \citeauthoryear {%
Ji%
\ \BBA {} Geroliminis%
}{%
Ji%
\ \BBA {} Geroliminis%
}{%
{\protect \APACyear {2012}}%
}]{%
Ji2012}
\APACinsertmetastar {%
Ji2012}%
\begin{APACrefauthors}%
Ji, Y.%
\BCBT {}\ \BBA {} Geroliminis, N.%
\end{APACrefauthors}%
\unskip\
\newblock
\APACrefYearMonthDay{2012}{}{}.
\newblock
{\BBOQ}\APACrefatitle {On the spatial partitioning of urban transportation
  networks} {On the spatial partitioning of urban transportation
  networks}.{\BBCQ}
\newblock
\APACjournalVolNumPages{Transportation Research Part B:
  Methodological}{46}{}{1639–1656}.
\PrintBackRefs{\CurrentBib}

\bibitem [\protect \citeauthoryear {%
Kirchhoff%
}{%
Kirchhoff%
}{%
{\protect \APACyear {1847}}%
}]{%
Kirchhoff1847}
\APACinsertmetastar {%
Kirchhoff1847}%
\begin{APACrefauthors}%
Kirchhoff, G.%
\end{APACrefauthors}%
\unskip\
\newblock
\APACrefYearMonthDay{1847}{}{}.
\newblock
{\BBOQ}\APACrefatitle {Über die Auflösung der Gleichungen, auf welche man bei
  der Untersuchung der linearen Vertheilung galvanischer Ströme geführt wird}
  {Über die auflösung der gleichungen, auf welche man bei der untersuchung
  der linearen vertheilung galvanischer ströme geführt wird}.{\BBCQ}
\newblock
\APACjournalVolNumPages{Annalen der Physik}{148}{}{497--508}.
\newblock
\doi{10.1002/andp.18471481202}
\PrintBackRefs{\CurrentBib}

\bibitem [\protect \citeauthoryear {%
Lebart%
}{%
Lebart%
}{%
{\protect \APACyear {1978}}%
}]{%
Lebart78}
\APACinsertmetastar {%
Lebart78}%
\begin{APACrefauthors}%
Lebart, L.%
\end{APACrefauthors}%
\unskip\
\newblock
\APACrefYearMonthDay{1978}{}{}.
\newblock
{\BBOQ}\APACrefatitle {Programme d'agr\'egation avec contraintes} {Programme
  d'agr\'egation avec contraintes}.{\BBCQ}
\newblock
\APACjournalVolNumPages{Cahiers de l'analyse des donn\'ees}{3}{3}{275--287}.
\newblock
\begin{APACrefURL} \url{http://www.numdam.org/item/CAD_1978__3_3_275_0/}
  \end{APACrefURL}
\PrintBackRefs{\CurrentBib}

\bibitem [\protect \citeauthoryear {%
Masser%
\ \BBA {} Brown%
}{%
Masser%
\ \BBA {} Brown%
}{%
{\protect \APACyear {1975}}%
}]{%
Masser75}
\APACinsertmetastar {%
Masser75}%
\begin{APACrefauthors}%
Masser, I.%
\BCBT {}\ \BBA {} Brown, P\BPBI J\BPBI B.%
\end{APACrefauthors}%
\unskip\
\newblock
\APACrefYearMonthDay{1975}{}{}.
\newblock
{\BBOQ}\APACrefatitle {Hierarchical aggregation procedures for interaction
  data} {Hierarchical aggregation procedures for interaction data}.{\BBCQ}
\newblock
\APACjournalVolNumPages{Environment and Planning A}{7}{}{509–523}.
\PrintBackRefs{\CurrentBib}

\bibitem [\protect \citeauthoryear {%
Murtagh%
}{%
Murtagh%
}{%
{\protect \APACyear {1985}}%
}]{%
Murtagh85}
\APACinsertmetastar {%
Murtagh85}%
\begin{APACrefauthors}%
Murtagh, F.%
\end{APACrefauthors}%
\unskip\
\newblock
\APACrefYearMonthDay{1985}{}{}.
\newblock
{\BBOQ}\APACrefatitle {A Survey of Algorithms for Contiguity-Constrained
  Clustering and Related Problems} {A survey of algorithms for
  contiguity-constrained clustering and related problems}.{\BBCQ}
\newblock
\APACjournalVolNumPages{The Computer Journal}{28}{}{82–88}.
\PrintBackRefs{\CurrentBib}

\bibitem [\protect \citeauthoryear {%
Murtagh%
\ \BBA {} Contreras%
}{%
Murtagh%
\ \BBA {} Contreras%
}{%
{\protect \APACyear {2012}}%
}]{%
Murtagh2012}
\APACinsertmetastar {%
Murtagh2012}%
\begin{APACrefauthors}%
Murtagh, F.%
\BCBT {}\ \BBA {} Contreras, P.%
\end{APACrefauthors}%
\unskip\
\newblock
\APACrefYearMonthDay{2012}{}{}.
\newblock
{\BBOQ}\APACrefatitle {Algorithms for hierarchical clustering: an overview}
  {Algorithms for hierarchical clustering: an overview}.{\BBCQ}
\newblock
\APACjournalVolNumPages{WIREs Data Mining and Knowledge
  Discovery}{2}{1}{86-97}.
\newblock
\begin{APACrefURL}
  \url{https://wires.onlinelibrary.wiley.com/doi/abs/10.1002/widm.53}
  \end{APACrefURL}
\newblock
\doi{https://doi.org/10.1002/widm.53}
\PrintBackRefs{\CurrentBib}

\bibitem [\protect \citeauthoryear {%
Openshaw%
}{%
Openshaw%
}{%
{\protect \APACyear {1977}}%
}]{%
Openshaw77}
\APACinsertmetastar {%
Openshaw77}%
\begin{APACrefauthors}%
Openshaw, S.%
\end{APACrefauthors}%
\unskip\
\newblock
\APACrefYearMonthDay{1977}{}{}.
\newblock
{\BBOQ}\APACrefatitle {A Geographical Solution to Scale and Aggregation
  Problems in Region-Building, Partitioning and Spatial Modeling} {A
  geographical solution to scale and aggregation problems in region-building,
  partitioning and spatial modeling}.{\BBCQ}
\newblock
\APACjournalVolNumPages{Transactions of the Institute of British
  Geographers}{2}{}{459–72}.
\PrintBackRefs{\CurrentBib}

\bibitem [\protect \citeauthoryear {%
Page%
\ \BBA {} Quintana%
}{%
Page%
\ \BBA {} Quintana%
}{%
{\protect \APACyear {2016}}%
}]{%
Garrit2016}
\APACinsertmetastar {%
Garrit2016}%
\begin{APACrefauthors}%
Page, G\BPBI L.%
\BCBT {}\ \BBA {} Quintana, F\BPBI A.%
\end{APACrefauthors}%
\unskip\
\newblock
\APACrefYearMonthDay{2016}{}{}.
\newblock
{\BBOQ}\APACrefatitle {{Spatial Product Partition Models}} {{Spatial Product
  Partition Models}}.{\BBCQ}
\newblock
\APACjournalVolNumPages{Bayesian Analysis}{11}{1}{265 -- 298}.
\newblock
\begin{APACrefURL} \url{https://doi.org/10.1214/15-BA971} \end{APACrefURL}
\newblock
\doi{10.1214/15-BA971}
\PrintBackRefs{\CurrentBib}

\bibitem [\protect \citeauthoryear {%
Pebesma%
}{%
Pebesma%
}{%
{\protect \APACyear {2018}}%
}]{%
Pebesma2018}
\APACinsertmetastar {%
Pebesma2018}%
\begin{APACrefauthors}%
Pebesma, E.%
\end{APACrefauthors}%
\unskip\
\newblock
\APACrefYearMonthDay{2018}{}{}.
\newblock
{\BBOQ}\APACrefatitle {{Simple Features for R: Standardized Support for Spatial
  Vector Data}} {{Simple Features for R: Standardized Support for Spatial
  Vector Data}}.{\BBCQ}
\newblock
\APACjournalVolNumPages{{The R Journal}}{10}{1}{439--446}.
\newblock
\begin{APACrefURL} \url{https://doi.org/10.32614/RJ-2018-009} \end{APACrefURL}
\newblock
\doi{10.32614/RJ-2018-009}
\PrintBackRefs{\CurrentBib}

\bibitem [\protect \citeauthoryear {%
Peixoto%
}{%
Peixoto%
}{%
{\protect \APACyear {2019}}%
}]{%
Peixoto19}
\APACinsertmetastar {%
Peixoto19}%
\begin{APACrefauthors}%
Peixoto, T\BPBI P.%
\end{APACrefauthors}%
\unskip\
\newblock
\APACrefYearMonthDay{2019}{}{}.
\newblock
{\BBOQ}\APACrefatitle {Bayesian Stochastic Blockmodeling} {Bayesian stochastic
  blockmodeling}.{\BBCQ}
\newblock
\BIn{} \APACrefbtitle {Advances in Network Clustering and Blockmodeling}
  {Advances in network clustering and blockmodeling}\ (\BPG~289-332).
\newblock
\APACaddressPublisher{}{John Wiley \& Sons, Ltd}.
\newblock
\begin{APACrefURL}
  \url{https://onlinelibrary.wiley.com/doi/abs/10.1002/9781119483298.ch11}
  \end{APACrefURL}
\newblock
\doi{https://doi.org/10.1002/9781119483298.ch11}
\PrintBackRefs{\CurrentBib}

\bibitem [\protect \citeauthoryear {%
{R Core Team}%
}{%
{R Core Team}%
}{%
{\protect \APACyear {2019}}%
}]{%
Rcore}
\APACinsertmetastar {%
Rcore}%
\begin{APACrefauthors}%
{R Core Team}.%
\end{APACrefauthors}%
\unskip\
\newblock
\APACrefYearMonthDay{2019}{}{}.
\newblock
{\BBOQ}\APACrefatitle {R: A Language and Environment for Statistical Computing}
  {R: A language and environment for statistical computing}{\BBCQ}\
  [\bibcomputersoftwaremanual].
\newblock
\APACaddressPublisher{Vienna, Austria}{}.
\newblock
\begin{APACrefURL} \url{https://www.R-project.org/} \end{APACrefURL}
\PrintBackRefs{\CurrentBib}

\bibitem [\protect \citeauthoryear {%
Randriamihamison%
, Vialaneix%
\BCBL {}\ \BBA {} Neuvial%
}{%
Randriamihamison%
\ \protect \BOthers {.}}{%
{\protect \APACyear {2020}}%
}]{%
Randriamihamison20}
\APACinsertmetastar {%
Randriamihamison20}%
\begin{APACrefauthors}%
Randriamihamison, N.%
, Vialaneix, N.%
\BCBL {}\ \BBA {} Neuvial, P.%
\end{APACrefauthors}%
\unskip\
\newblock
\APACrefYearMonthDay{2020}{}{}.
\newblock
{\BBOQ}\APACrefatitle {{Applicability and Interpretability of Ward Hierarchical
  Agglomerative Clustering With or Without Contiguity Constraints}}
  {{Applicability and Interpretability of Ward Hierarchical Agglomerative
  Clustering With or Without Contiguity Constraints}}.{\BBCQ}
\newblock
\APACjournalVolNumPages{Journal of Classification}{}{}{}.
\newblock
\begin{APACrefURL} \url{https://hal.archives-ouvertes.fr/hal-02294847}
  \end{APACrefURL}
\newblock
\doi{https://dx.doi.org/10.1007/s00357-020-09377-y}
\PrintBackRefs{\CurrentBib}

\bibitem [\protect \citeauthoryear {%
Rasmussen%
\ \BBA {} Ghahramani%
}{%
Rasmussen%
\ \BBA {} Ghahramani%
}{%
{\protect \APACyear {2001}}%
}]{%
Rasmussen01}
\APACinsertmetastar {%
Rasmussen01}%
\begin{APACrefauthors}%
Rasmussen, C.%
\BCBT {}\ \BBA {} Ghahramani, Z.%
\end{APACrefauthors}%
\unskip\
\newblock
\APACrefYearMonthDay{2001}{}{}.
\newblock
{\BBOQ}\APACrefatitle {Infinite Mixtures of Gaussian Process Experts} {Infinite
  mixtures of gaussian process experts}.{\BBCQ}
\newblock
\BIn{} T.~Dietterich, S.~Becker\BCBL {}\ \BBA {} Z.~Ghahramani\ (\BEDS),
  (\BVOL~14).
\newblock
\APACaddressPublisher{}{MIT Press}.
\newblock
\begin{APACrefURL}
  \url{https://proceedings.neurips.cc/paper/2001/file/9afefc52942cb83c7c1f14b2139b09ba-Paper.pdf}
  \end{APACrefURL}
\PrintBackRefs{\CurrentBib}

\bibitem [\protect \citeauthoryear {%
Schwaller%
\ \BBA {} Robin%
}{%
Schwaller%
\ \BBA {} Robin%
}{%
{\protect \APACyear {2017}}%
}]{%
Schwaller17}
\APACinsertmetastar {%
Schwaller17}%
\begin{APACrefauthors}%
Schwaller, L.%
\BCBT {}\ \BBA {} Robin, S.%
\end{APACrefauthors}%
\unskip\
\newblock
\APACrefYearMonthDay{2017}{}{}.
\newblock
{\BBOQ}\APACrefatitle {Exact Bayesian inference for off-line change-point
  detection in tree-structured graphical models} {Exact bayesian inference for
  off-line change-point detection in tree-structured graphical models}.{\BBCQ}
\newblock
\APACjournalVolNumPages{Statistics and Computing}{27}{}{1331--1345}.
\newblock
\doi{10.1007/s11222-016-9689-3}
\PrintBackRefs{\CurrentBib}

\bibitem [\protect \citeauthoryear {%
Scrucca%
, Fop%
, Murphy%
\BCBL {}\ \BBA {} Raftery%
}{%
Scrucca%
\ \protect \BOthers {.}}{%
{\protect \APACyear {2016}}%
}]{%
Scrucca16}
\APACinsertmetastar {%
Scrucca16}%
\begin{APACrefauthors}%
Scrucca, L.%
, Fop, M.%
, Murphy, T\BPBI B.%
\BCBL {}\ \BBA {} Raftery, A\BPBI E.%
\end{APACrefauthors}%
\unskip\
\newblock
\APACrefYearMonthDay{2016}{}{}.
\newblock
{\BBOQ}\APACrefatitle {{mclust} 5: clustering, classification and density
  estimation using {G}aussian finite mixture models} {{mclust} 5: clustering,
  classification and density estimation using {G}aussian finite mixture
  models}.{\BBCQ}
\newblock
\APACjournalVolNumPages{The {R} Journal}{8}{1}{289--317}.
\newblock
\begin{APACrefURL} \url{https://doi.org/10.32614/RJ-2016-021} \end{APACrefURL}
\PrintBackRefs{\CurrentBib}

\bibitem [\protect \citeauthoryear {%
Teixeira%
, Assun{\c{c}}{{\~a}}o%
\BCBL {}\ \BBA {} Loschi%
}{%
Teixeira%
\ \protect \BOthers {.}}{%
{\protect \APACyear {2019}}%
}]{%
Leonardo2019}
\APACinsertmetastar {%
Leonardo2019}%
\begin{APACrefauthors}%
Teixeira, L\BPBI V.%
, Assun{\c{c}}{{\~a}}o, R\BPBI M.%
\BCBL {}\ \BBA {} Loschi, R\BPBI H.%
\end{APACrefauthors}%
\unskip\
\newblock
\APACrefYearMonthDay{2019}{}{}.
\newblock
{\BBOQ}\APACrefatitle {Bayesian Space-Time Partitioning by Sampling and Pruning
  Spanning Trees} {Bayesian space-time partitioning by sampling and pruning
  spanning trees}.{\BBCQ}
\newblock
\APACjournalVolNumPages{Journal of Machine Learning Research}{20}{85}{1--35}.
\newblock
\begin{APACrefURL} \url{http://jmlr.org/papers/v20/16-615.html}
  \end{APACrefURL}
\PrintBackRefs{\CurrentBib}

\bibitem [\protect \citeauthoryear {%
Traag%
, Waltman%
\BCBL {}\ \BBA {} Van~Eck%
}{%
Traag%
\ \protect \BOthers {.}}{%
{\protect \APACyear {2019}}%
}]{%
Traag19}
\APACinsertmetastar {%
Traag19}%
\begin{APACrefauthors}%
Traag, V\BPBI A.%
, Waltman, L.%
\BCBL {}\ \BBA {} Van~Eck, N\BPBI J.%
\end{APACrefauthors}%
\unskip\
\newblock
\APACrefYearMonthDay{2019}{}{}.
\newblock
{\BBOQ}\APACrefatitle {From Louvain to Leiden: guaranteeing well-connected
  communities} {From louvain to leiden: guaranteeing well-connected
  communities}.{\BBCQ}
\newblock
\APACjournalVolNumPages{Scientific reports}{9}{1}{1--12}.
\PrintBackRefs{\CurrentBib}

\end{thebibliography}

\appendix

\end{document}